\documentclass[conference]{IEEEtran}
\IEEEoverridecommandlockouts
\usepackage{cite}
\usepackage{amsmath,amssymb,amsfonts}
\usepackage{algorithmic}
\usepackage{graphicx}
\usepackage{textcomp}
\usepackage{xcolor}
\usepackage{caption}
\usepackage{multirow}
\usepackage{subcaption}
\usepackage{xcolor,colortbl}
\usepackage{makecell}
\usepackage{booktabs}
\def\BibTeX{{\rm B\kern-.05em{\sc i\kern-.025em b}\kern-.08em
    T\kern-.1667em\lower.7ex\hbox{E}\kern-.125emX}}
\begin{document}

\title{Developers' Visuo-spatial Mental Model and Program Comprehension}

\author{\IEEEauthorblockN{Abir Bouraffa}
\IEEEauthorblockA{\textit{Applied Software Technology} \\
\textit{Universität Hamburg}\\
Hamburg, Germany \\
abir.bouraffa@uni-hamburg.de}
\and
\IEEEauthorblockN{Gian-Luca Fuhrmann}
\IEEEauthorblockA{\textit{Applied Software Technology} \\
\textit{Universität Hamburg}\\
Hamburg, Germany 
}
\and
\IEEEauthorblockN{Walid Maalej}
\IEEEauthorblockA{\textit{Applied Software Technology} \\
\textit{Universität Hamburg}\\
Hamburg, Germany \\
walid.maalej@uni-hamburg.de
}
}
\maketitle

\begin{abstract}
Previous works from research and industry have proposed a spatial representation of code in a canvas, arguing that  a navigational code space  confers developers the freedom to organise elements according to their understanding.
By allowing developers to translate logical relatedness into spatial proximity, this code representation could aid in code navigation and comprehension.
However, the association between developers' code comprehension and their visuo-spatial mental model of the code is not yet well understood.
This mental model is affected on the one hand by the spatial code representation and on the other by the visuo-spatial working memory of developers.

We address this knowledge gap by conducting an online experiment with 20 developers following a between-subject design.
The control group used a conventional tab-based code visualization, while the experimental group used a code canvas to complete three code comprehension tasks.
Furthermore, we measure the participants' visuo-spatial working memory using a Corsi Block test at the end of the tasks. 
Our results suggest that, overall, neither the spatial  representation of code nor the visuo-spatial working memory  of developers has a significant impact on comprehension performance. 
However, we identified significant differences in the time dedicated to different comprehension activities such as navigation, annotation, and UI interactions. 

\end{abstract}


\section{Introduction}

Developers spend a significant amount of their work time on code comprehension. 
Previous research has established that this activity covers as high as 70\% of a developer's time \cite{Minelli2014}, whether they are debugging, reviewing code, or trying to add a new feature \cite{Maalej:TOSEM:14}. 
Research on program comprehension has traditionally followed two major goals \cite{SiegmundCC2016}. The first is to understand developers' mental models of code \cite{siegmund2017measuring, peitek2021program}, their behaviour, and the factors influencing comprehension performance \cite{hofmeister2017shorter, Peitek:FSE:2022}.  The second is to design tools that assist developers, particularly by visualizing and aiding  navigation \cite{kersten2006using}. 
These research efforts have helped shape today's IDEs, which have come a long way in providing a better development experience.
Nevertheless, the bento-box design (whereby the main IDE window is divided  into rectangular areas containing file trees, editors, navigators, the console, etc.~\cite{DeLineCodeCanvas2010}) has remained for the most part unrivalled despite proposed alternatives\cite{BragdonCodeBubbles2010}\cite{DeLineCollab2012}\cite{DeLineCodeCanvas2010}\cite{Patch2014} that failed to make it to the mainstream.

In studying program comprehension, due attention has been given to the effect of several confounding factors.
However, a literature survey of the field by Siegmund et al.~\cite{Siegmund2015Confounders} found that programming experience, familiarity with the tools, and familiarity with the study object are the most measured confounders.
Individual characteristics such as intelligence were found to be rarely included in such studies.
Yet, a cognitive activity such as program comprehension is likely affected by the cognitive abilities of developers.

This work focuses on developers' visuo-spatial mental model, a rarely studied aspect of code understanding, and its association with comprehension performance. 
We study this association from two perspectives: first by investigating the effect of using  alternative visual representations of code on comprehension; and second, by taking the visuo-spatial working memory of developers into account.
Visuo-spatial working memory is a cognitive characteristic representing a natural limitation to the visual information load a human brain can hold and process in the short-term \cite{Bancifra22VisuospatialMemory}.

We report on an empirical study with 20 developers, who performed three comprehension tasks in an online setting lasting approximately one hour. 
The control group used a tab-based code visualization tool to solve the tasks, while  the experimental group used a code canvas. 
We measured the time to completion, the accuracy of the code localization and annotation, as well as the visuo-spatial working memory with a Corsi Block test. 
Our main motivation is first to assess whether the freedom of arrangement conferred by a code canvas has any influence on code comprehension performance and on the behaviour of developers in the IDE.
Secondly, we would like to measure the association these two aspects have with the cognitive trait of visuo-spatial working memory.
Our research questions are thus formulated as follows:
\begin{itemize}
\item RQ1: Does the \textbf{spatial representation} of code influence code comprehension performance?
\item RQ2: How does the use of a code canvas influence the time allocated to specific \textbf{comprehension activities} such as navigation and understanding?
\item RQ3: How is developer's \textbf{visuo-spatial working memory} associated with code comprehension performance?
\end{itemize}

We did not observe any significant effect of code representation and visuo-spatial working memory on comprehension performance. 
Overall, while participants in the canvas group needed less time to complete the experiment tasks, the annotation  accuracy in the control group was higher.  
We did not find significant differences in the time allocated to different development activities between the two groups.
However, we found a significant negative correlation between visuo-spatial working memory and time spent on annotating and UI interactions as well as a positive correlation with time spent on navigation.
When analyzing the canvas trace of the experimental group, we found that participants with higher visuo-spatial memory performance tended to use complex canvases with many files, while the rest tended to ``clean up'' the canvas and arrange the files based on logical relatedness. 

The remainder of the paper is structured as follows. Section \ref{sec:rel_work} introduces the foundation and discusses related work. 
Section \ref{sec:design} presents our experimental design, including the code visualization tool, the comprehension tasks, as well as the experiment participants, procedures, and measured variables.
Section \ref{sec:results} summarises the results according to the research questions.
Finally, Section \ref{sec:discussion} discusses the implications of our work for research and practice, while Section \ref{sec:threats} summarises the threats to validity and Section \ref{sec:conclusion} concludes the paper.

\section{Related Work}
\label{sec:rel_work}
\subsection{Spatial Code Representation}
The work of Bragdon et al.~\cite{BragdonCodeBubbles2010} on Code Bubbles counts as one of the earliest proposals for an alternative to the tab-based IDE design.
The authors propose a representation of code fragments such as methods or function documentation as bubbles in a large virtual space, where they can be clustered into working sets.
Similarly, the  \textit{Code Canvas} tool developed by DeLine et al.~\cite{DeLineCodeCanvas2010} offers a zoomable surface to  accommodate project artefacts, which can be organised into layers of information.
The authors of both works later collaborated jointly with a team from Microsoft Visual Studio to develop an industrial version of the Code Bubbles paradigm  called \textit{Debugger Canvas} \cite{DeLineCollab2012}.
Prior to this project, DeLine et al.~\cite{DeLineThumbnails2006} worked on using thumbnails to enhance intra-file navigation.
This thumbnail overview of the file has made its way to mainstream IDEs such as Visual Studio Code where it has been named the file minimap\footnote{https://code.visualstudio.com/docs/getstarted/userinterface\#\_minimap}.

More recently, Adeli et al.~\cite{Adeli2020} presented a between-subject study in which project newcomers were either assigned a conventional IDE (Eclipse) or a canvas-based IDE called Synectic allowing users to spatially arrange cards of relevant information (code, web pages, etc.), along with annotations that link information to individual cards.
The authors focused on the effects of annotations in supporting information foraging by linking relevant information and code on a canvas-based IDE. 
Participants using Synectic received a canvas pre-populated with cards containing onboarding documentation.
The control group using Eclipse received an onboarding document file that linked to the code.
The authors compared the two groups in terms of time needed to accomplish the tasks, the accuracy of responses, and self-reported cognitive load.
Their results showed that providing the right information at the right place and time helped newcomers answer comprehension questions with significantly more accuracy in less time.

Another research direction in the field of software visualization is inspired by the city metaphor.
The city metaphor and the notion of habitability described by Wettel et al.~convey the idea that a software system is a ``\textit{physical space with strong orientation points''}~\cite{Wettel07Habitability}.
 Wettel et al.~\cite{Wettel11ICSE} reported the results of a controlled experiment showing that a 3D software visualization, implemented in the CodeCity tool, lead to a significant increase in comprehension task correctness and a decrease in completion time.
\begin{figure*}[h!]
\centering
    \begin{subfigure}[b]{0.45\textwidth}
        \includegraphics[scale=0.39]{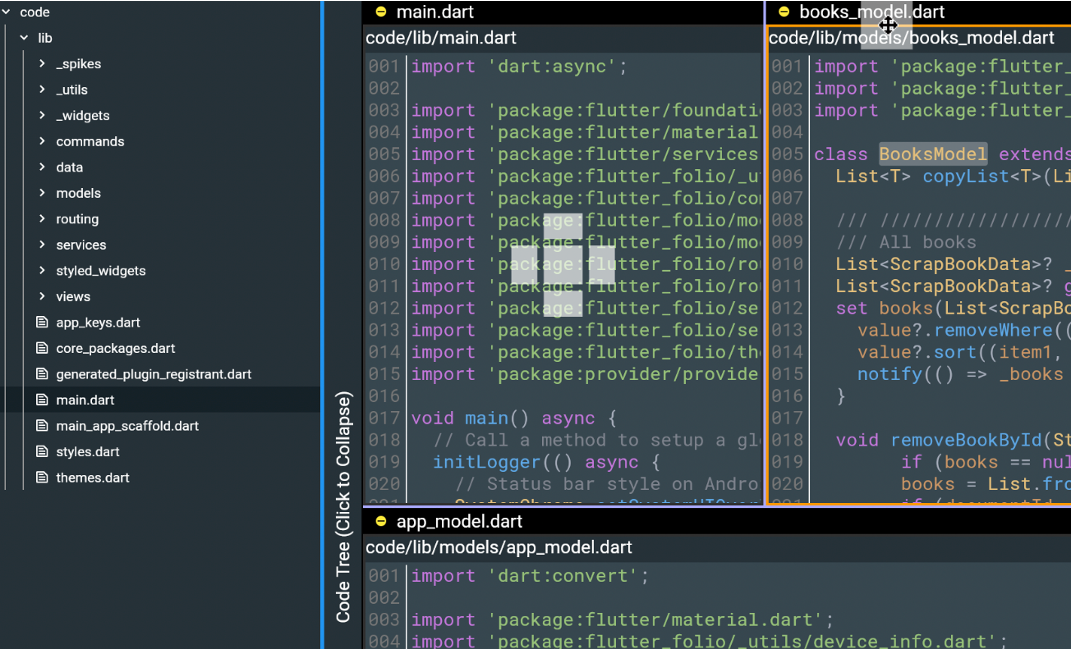}
        \caption{Tab-based code visualization}
    \label{fig:tab-based-tool}
    \end{subfigure}
    ~
    \begin{subfigure}[b]{0.45\textwidth}
        \includegraphics[scale=0.4]{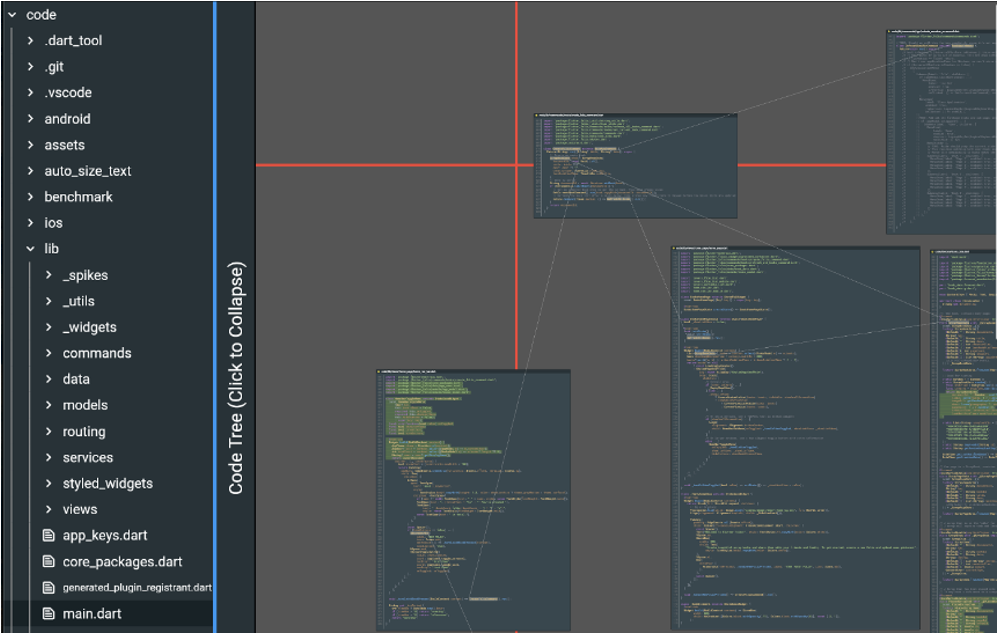}
        \caption{Canvas-based code visualization}
    \label{fig:canvas-based-tool}
    \end{subfigure}
    \caption{Tab-based vs. canvas-based code visualization in the experimental tool.}
    \label{fig:tool-code-visualisation}
\end{figure*}

\subsection{Cognitive Maps}
Cognitive maps are a concept proposed by Tolman \cite{Tolman1948CMaps} describing an internal representation of an external environment used for navigation purposes.
Spatial information can be described as either egocentric, i.e.~relative to the person, or allocentric, i.e.~relative to the environment \cite{Burgess2002SMemory}.
Mental models built up during program comprehension have been repeatedly theorised to take the form of a cognitive map within spatial memory.
For instance, Cox et al.~\cite{Cox2005CMap} suggested that during program comprehension a cognitive map is generated, which is referred to as a code-to-concept map allowing the programmer to navigate software.
Green and Navarro \cite{Green1995Imagery} moreover found that program comprehension is facilitated when related program concepts are in close physical proximity.

In a study by Jones et al.~\cite{Jones2007SpatialSkills}, students with higher spatial ability, evidenced by their score on a mental rotation test, were found to need a shorter time to complete a code comprehension task.
Additionally, the results showed they spent less time searching through the code but jumped more frequently within and between class files.
Other studies such as Huang et al.~\cite{Huang2019fmri} found  that core software engineering
tasks use visuo-spatial cognitive processes.
In fact, using medical imaging, the authors discovered that  similar parts of the brain are activated to solve mental rotation problems and tree-based data structure problems.
CS Education researchers have attempted to model these cognitive functional similarities, which resulted in a plethora of theories. 
For instance, Margulieux et al.~\cite{MargulieuxCognitive2019} proposed the \textit{Spatial Encoding Strategy (SpES)}, a theory behind the transfer-ability between spatial ability and programming, whereby developing spatial skills aided in acquiring generalisable strategies to form mental representations of non-verbal information~\cite{MargulieuxCognitive2019}.

\subsection{Visuo-spatial Working Memory}
Visuo-spatial working memory refers to the ability to temporarily store and mentally manipulate visual information \cite{Bancifra22VisuospatialMemory}. 
Visual processing has been investigated by researchers  as a perceptual process, a physiological process, and  a memory function \cite{Bancifra22VisuospatialMemory}.
According to Duff et al., ~\cite{Duff99VisuoSpatialWM}, visuo-spatial working memory involves the usage of multiple cognitive components including a visual cache for the temporary storage of the appearance of objects and an inner scribe for rehearsing targeted movements between locations in the environment.

A common  test to measure visuo-spatial working memory is the Corsi Block test. 
This test is commonly used by clinical neuropsychologists as well as developmental and cognitive psychologists to study learning disorders as well as other conditions such as Alzheimer’s and Huntington’s disease.
The Corsi block test is also used on healthy subjects to measure their visuo-spatial abilities \cite{Furley10CorsiBlockAthletes}.
Rafaella et al.~\cite{Raffaella09Corsi} for instance used the Corsi block test to explore the relationship between individual differences in visuo-spatial working memory and wayfinding performance.
The test has also been applied in human-centred computing  by Jin et al. \cite{Jin18CORSI_UI} to investigate how individual traits influence the perception of visual user interfaces for diversity-enhanced music recommender systems.
 Huang and Klippel \cite{Huang20VisualRealism} applied the test to investigate the effects of visual realism on object location memory in a VR setting taking into account  spatial ability.

Originally developed by Philip M. Corsi \cite{Corsi1972HumanMA}, the Corsi Block-Tapping test is comparable to the digit span test for assessing verbal short-term memory~\cite{OstroskyDigitSpan2007}.
In the original version of the test, the examiner uses nine cubes mounted on a board and taps a sequence of blocks which the subject has to repeat subsequently in the correct sequential order. 
By increasing the length of the sequence, the capacity of the visuo-spatial short-term memory can be measured.

\section{Study Design}
\label{sec:design}

\begin{figure*}[!h]
\begin{minipage}[c]{0.35\textwidth}
\includegraphics[scale=0.28]{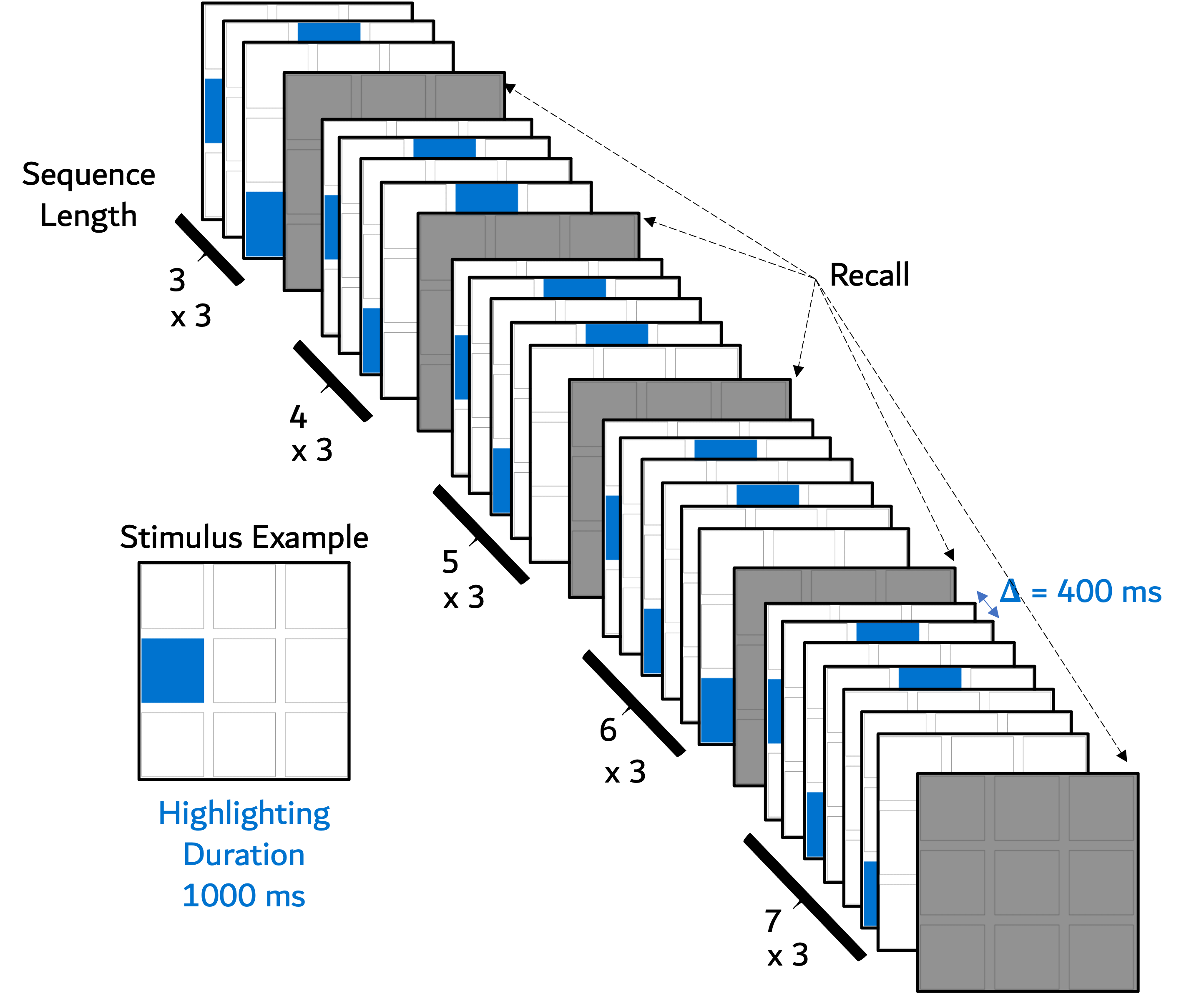}
    \caption{Corsi Block test.}
    \label{fig:corsi_block_test}
\end{minipage}
\hfill
\begin{minipage}[c]{0.67\textwidth}
    \centering
    \includegraphics[scale=0.19]{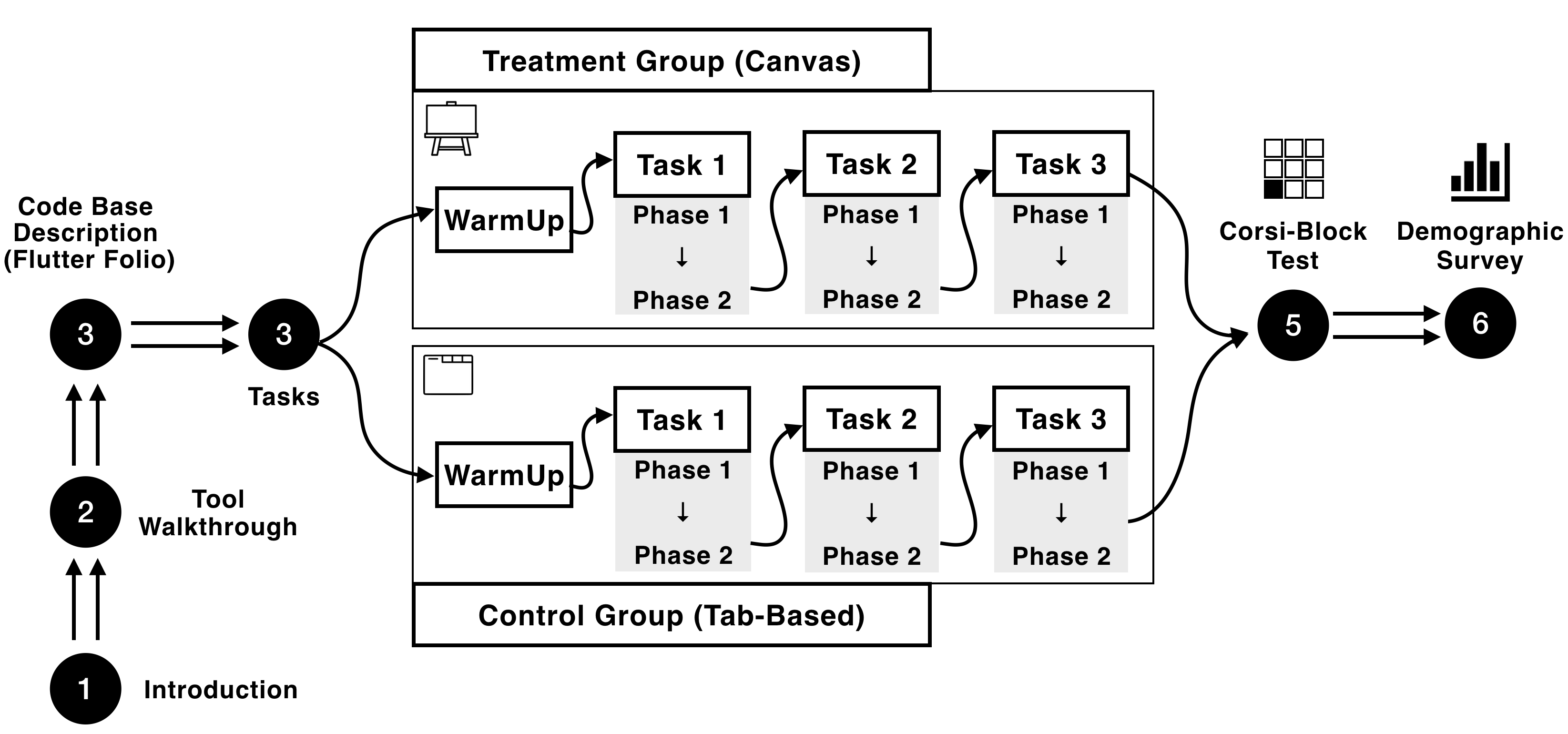}
    \caption{Overview of the experiment design.}
    \label{fig:experiment_overview}
\end{minipage}

\end{figure*}

To answer our research questions, we conducted an online between-subject experiment as outlined in Figure \ref{fig:experiment_overview}. 
We asked participants to complete three code comprehension tasks each consisting of a navigation and a comprehension phase.
Participants were randomly assigned to the control and treatment groups.
We developed two versions of a  web-based code visualization tool: a conventional tab-based version which we made available to participants of the control group and a canvas-based version used by participants of the treatment group.
After finishing the three tasks, participants were asked to take an online Corsi Block test and to answer a short demographic survey.

\subsection{Code Visualization}

To control the experimental  environment (i.e. , IDE features) and to avoid user experience bias, we developed a web tool to visualise a code base in two versions: a canvas-based version and a tab-based version.
Figure \ref{fig:tool-code-visualisation} shows the code representation in both versions.
A primary sidebar showing the project's file tree is provided in both versions.
In the canvas version (in Figure \ref{fig:canvas-based-tool}), the source code files are represented as nodes inside a zoomable canvas, allowing the user to open multiple files at once and to arrange them freely by dragging them to the desired location. 
In the tab-based version (in Figure \ref{fig:tab-based-tool}), the user can open multiple files with the latest opened tab being the main focus of the screen.
As in common IDEs, the screen can be split between multiple open tabs.

By design, we disabled code editing in both versions. 
Instead, the user is able to highlight code sections using pointer drags, similar to text highlighting in a browser.
Upon releasing the mouse selection, a highlighting overlay of the selected area is created.
We use different highlighting colours for each of the tasks to enable participants to distinguish their solutions for each task.
We use semantic tokenization through calls to the  Language Server Protocol (LSP) to support syntax highlighting, \textit{Go to Definition}, and \textit{Find all References} look-ups on source code tokens.
Additionally, a toggle button allows the user to visualise or hide the link between the source and the target tokens when these lookups are made in the canvas version.

During each task, a task description is shown inside a minimizable window at the top right corner of the screen.
Annotations to the source code  appear as a list below the task description so that participants can review and remove them if necessary.
During the experiment, we logged the following  interactions with the tool:
\begin{itemize}
    \item Open a file from the file tree
    \item Perform a lookup via \textit{Go to Definition}
    \item Perform a lookup via \textit{Find all References}
    \item Highlight a code section (referred to as an annotation)
    \item Delete a highlighting 
    \item Move a file
    \item Hide a file (equivalent to minimise a file)
\end{itemize}

\subsection{Corsi Block Test}
To assess the visuo-spatial working memory of the participants, we use a web version of the Corsi Block test taken from the Tatool  test battery\footnote{https://github.com/tatool/tatool-web}.
Tatool is an open-source framework for psychological studies originally developed in Java by von Bastian et al.~\cite{Von_Bastian2013Tatool} and later released as a web version.
The default distribution of the Corsi Block test (available in Tatool\footnote{http://www.tatool-web.com/\#!/doc/lib-exp-Corsi Block.html}) displays a 3x3 matrix of white blocks.
In each iteration of the test, matrix blocks are highlighted one after another.
Each single block in a sequence remains highlighted for a duration of $1000 ms$.
Subsequently, a delay of $400 ms$ is configured before the next block highlighting starts as described in Figure \ref{fig:corsi_block_test}.
This corresponds to the default configuration of the cognitive test battery.
The total number of highlighted blocks in an iteration corresponds to the current sequence length.
After each iteration, a participant is asked to recall the positions of the highlighted blocks in the correct serial order by clicking on the respective cells in the matrix.
We configured the test to have a sequence length between 3 and 7 with three iterations for each sequence.
Thus, the maximum score that a participant can achieve on the test is $75 = 3*(3+4+5+6+7)$.
Prior to the actual test, participants were given a short description of the test and were asked to perform a dry run to familiarise themselves with the task.

\begin{table*}[!h]
    \centering
     \caption{Overview of comprehension tasks performed in the experiment.}
    \renewcommand{\arraystretch}{1}
    \begin{tabular}{c c l}
    \toprule
        Task &  Phase & Description\\
        \midrule
        \multirow{4}{*}{1} & 0 & \makecell[l]{\textbf{One of the app’s features in the Desktop version allows the user to share a link to a book via a button.} \\Please trace the program flow from when the corresponding button is pressed to when the link is made available.
        }\\ 
        \\
        & 1 & \makecell[l]{A user complained that the shared link contains their email address in plain text. \\Your task is to find and mark the code line(s) that resulted in this program bug. \\ Before clicking “continue”, please highlight the involved line(s) across the involved file(s).}\\ 
        \midrule
        \multirow{5}{*}{2} & 0 & \makecell[l]{\textbf{When the user deletes a page of a book (via a button), the next
page in the book is shown as the current page.}\\ Please trace the program flow from when the corresponding button is pressed to when the next page is shown. 
}\\ 
        \\
        & 1 & \makecell[l]{A user complained that an error occurred citing \textit{“delete\_page\_command.dart”} at \textit{“line 33”}, when they deleted the only page in a book.\\ Your task is to find the corresponding code line(s) that caused the bug. Before clicking “continue”, please highlight the involved line(s)\\ across the involved file(s). 
        }\\
        \midrule
        \multirow{5}{*}{3} & 0 & \makecell[l]{\textbf{The App uses a firebase service to upload data into cloud firestore.}\\ This service is implemented in two classes: \textit{“NativeFirebaseService”} and \textit{“DartFirebaseService”}.\\ Which version of the App (desktop, mobile, web) uses which service class? Find the location where the decision is made to use\\ either one of the two services. Before clicking “continue”, please highlight the involved line(s) across the involved file(s). 
        }\\ 
        \\
        & 1 & \makecell[l]{How does the call to the firestore API differ between the two classes \textit{“NativeFirebaseService”} \\and \textit{“DartFirebaseService”} when adding a new document? Please annotate the differing code line in both classes.}\\
        \bottomrule
    \end{tabular}
   
    \label{tab:tasks}
\end{table*}

\subsection{Project and Tasks}
As a development project, we chose the open source scrap-booking app {Flutter Folio}\footnote{https://github.com/gskinnerTeam/flutter-folio}, which includes $\sim15,000$ lines of code. 
The project was created by gskinner to showcase best practices for developing multi-platform apps in Flutter. 
The app allows the user to create an image book called \textit{folio} including pages in which images can be uploaded and annotated by adding text and emojis called \textit{scraps}.
The user can share the scrapbook with others through a link-sharing feature.
The app is available on  web, mobile, and desktop. 
The web version is available online for demonstration purposes\footnote{https://www.flutterfolio.com/}.

We asked participants to complete three comprehension tasks detailed in Table \ref{tab:tasks}. 
We designed the tasks to last approximately  one hour in total.  
In each task, participants were asked in Phase 0 to find and annotate code lines relevant to a given app feature by highlighting them.
Then, in Phase 1, they were asked in-depth questions about the code section, which were formulated as debugging questions for tasks 1 and 2 and a comparison of two program behaviours in task 3.

In the first task, participants were asked to trace the link-sharing functionality starting from the click of the corresponding button, which was present in three views of the app.
The second part of this task asked the participants to locate the code section responsible for showing the user email in plain text within the shared link, for which we modified the original code accordingly.
The second task was taken from a bug that was fixed in the Flutter Folio repository\footnote{https://github.com/gskinnerTeam/flutter-folio/commit/0b0894885f0db48333\\57026e6ff83a40c3a1361d}.
The bug consisted of an incorrect resetting of the current page when one of the pages in a scrapbook is deleted.
Lastly, the third task asked the participants to determine which firestore service is used on different platforms (web, desktop, mobile) and how the API calls  differed  when a new document was added.

\subsection{Participants}
We recruited participants among  students who took part in an advanced app development project using Flutter/Dart in the past 3 years.
Additionally, we approached developers from our personal networks, who were familiar with Flutter and invited them to participate.
We had two recruitment rounds: in the  first round, we gathered 19 participations. 
In the second round, we managed to collect further 7 participations leading to a total of 26 participations.
After carefully investigating the collected data, we decided to eliminate 6 participations on the grounds of unreliable or incomplete data: two of these participants reported experiencing massive performance problems with our web tool, likely also due to a poor internet connection.
Two further participants did not complete multiple tasks (no annotations made), and one participant did not complete the Corsi block test and the subsequent survey.
One participant received the wrong tool link for the assigned group by mistake and had to be eliminated from the final dataset.
Finally, we had 20 complete participations, which we analyse in the remainder of the paper. 

All 20 participants except two had at least two years of experience with software development.
Among them, 13 are working as developers, one as analyst, one as an IT consultant, and 5 are still students. 
In total, 10 participants in our sample reported having 1 year or less of development experience in Dart/Flutter, 6 reported 2 years, 2 participants had 3 to 5 years of experience and two reported no prior experience with this framework.

The majority (12 participants) stated writing code at least once a day. 
Only three participants stated programming once a month while the rest stated writing code at least on a weekly basis.
When asked how many hours they have worked before taking the experiment, 5 participants stated not having worked at all, 11 participants had worked 1 to 8 hours while the rest worked  more than 8 hours for an overall mean of 4.65 hours.

Participants were allowed to take part in the online experiment at their convenience. 
However, interruptions of the experiment were strongly discouraged.
As a token of appreciation, we raffled two value vouchers among participants, each equivalent to 50 euros.

\begin{table}
\centering
    \caption{Variables measured in the experiment.}
    \label{tab:experiment_variables}
    \renewcommand{\arraystretch}{1}
    \begin{tabular}{ll}
    \toprule
    \textbf{Extraneous Variables} & \textbf{Type}\\
    
    Development experience &Ordinal/Interval\\
    Dart experience &Ordinal/Interval\\
    Programming frequency &Ordinal/Interval\\
    Working hours before the experiment &Ratio\\
    Mental fitness before the experiment &Ordinal\\
    \midrule
    \textbf{Measured Confounders} & \textbf{Type}\\
    Corsi Block test score &Ratio\\
    \midrule
    \textbf{Independent Variables } & \textbf{Type}\\
    Group &Nominal\\
    \midrule
    \textbf{Dependent Variables} & \textbf{Type }\\
    Net time &Ratio\\
    File accuracy &Ratio\\
    Annotation Accuracy & Ratio\\
    Understanding Time & Ratio\\
    Annotation Time & Ratio\\
    Navigation Time & Ratio\\
    UI Interaction Time &Ratio\\
    \bottomrule
\end{tabular}
\end{table}

\subsection{Experiment Variables}

Table \ref{tab:experiment_variables} outlines the variables used in this experiment.
The extraneous variables such as development experience and programming frequency were collected through the demographic survey while the spatial span and the number of correct trials stem from the Corsi Block test.
All dependent variables were extracted from the raw server logs resulting from the interactions with the tool.
We assigned an ordinal score between 1 and 5 to the answers to survey questions about development experience (\textit{none} to \textit{5-6 years}), experience with the Dart programming language, programming frequency (\textit{none} to \textit{daily})  and mental fitness at the start of the experiment (\textit{very tired} to \textit{very fit}).

\begin{table}[!ht]
    \centering
     \caption{Demographic characteristics of the participants.}
\label{tab:means_independent_variables}
    \begin{tabular}{llrrrrrrrrrrrrr}
    \toprule
         & ~ & \textbf{Prof. Dev. Experience}  & \textbf{Occupation} & \textbf{Prog. Frequency} \\
        \midrule
        \multirow{10}{*}{\rotatebox[origin=c]{90}{\textbf{Canvas}}}&
 P1&  3 to 5 years & developer & once a day  \\
&P2&  2 years &   developer & once a week\\
&P3& 3 to 5 years &  developer & once a day  \\
&P4&  no experience  &     student & once a day \\
&P5&  2 years  &     analyst & once a month  \\
&P6& 3 to 5 years &  developer & once a day  \\
&P7& 3 to 5 years  &     student &  once a month  \\
&P8&  3 to 5 years  &     student &once a day \\
&P9&  2 years &   developer & once a week \\
&P10& 3 to 5 years &   developer &  once a day \\
        \midrule
        \multirow{10}{*}{\rotatebox[origin=c]{90}{\textbf{Control}}}
&P11 &              3 5 years &  developer &              once a day \\
&P12  &              3 to 5 years &  developer &             once a week \\
&P13  &                2 years &  developer &             once a week \\
&P14  &                2 years &     student &              once a day \\
&P15  &                2 years &  developer &              once a day \\
&P16  &                2 years &  developer &              once a day \\
&P17  &                2 years &  developer &              once a day \\
&P18  &          no experience &     student &            once a month \\
&P19  &              3 to 5 years &  developer &              once a day \\
&P20  &             6 to 10 years &       IT consultant &             once a week \\
        \bottomrule
    \end{tabular}
   
\end{table}

\begin{figure*}[h]
    \centering
    \begin{subfigure}[b]{0.3\textwidth}
        \includegraphics[scale=0.38]{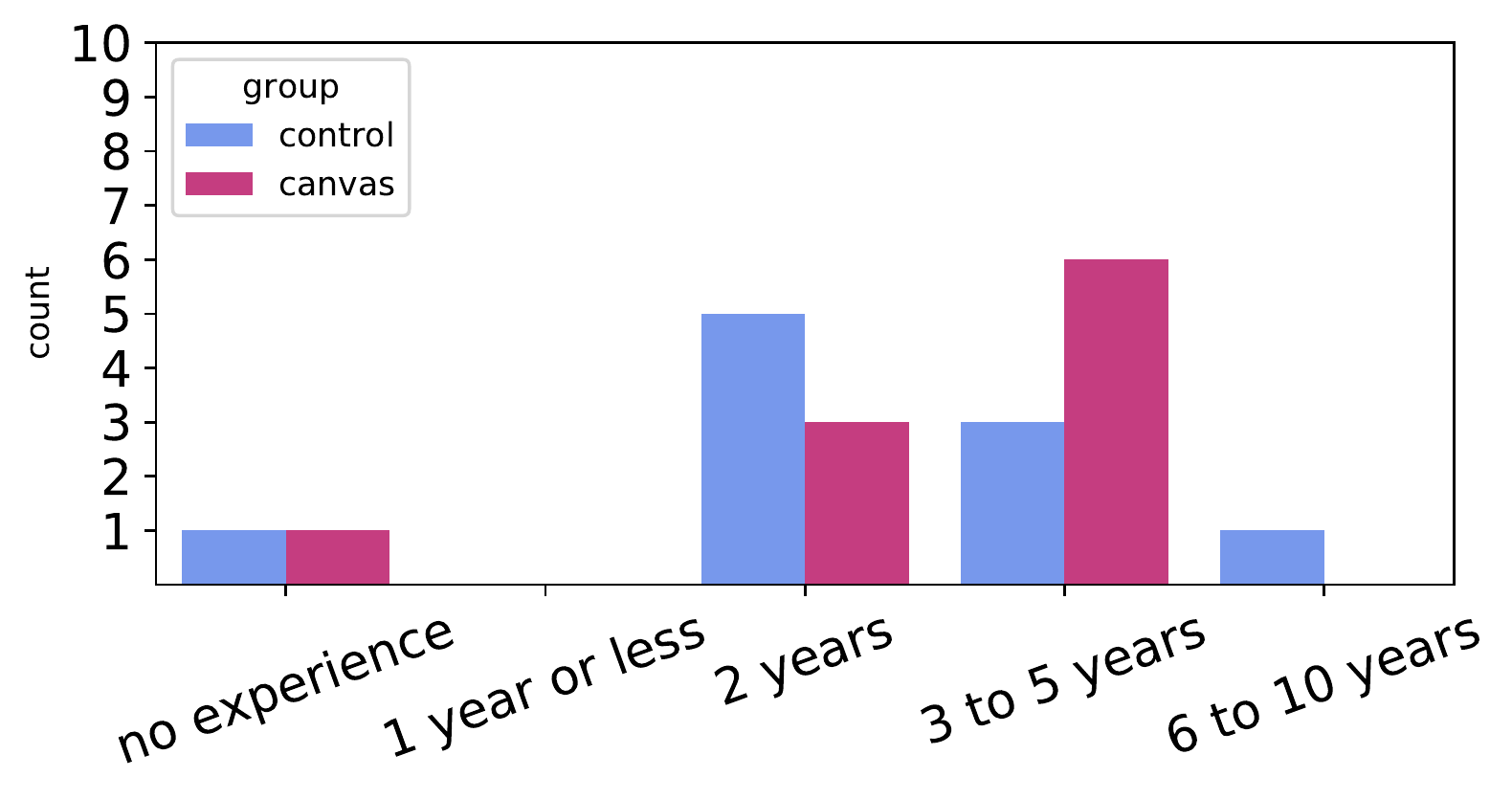}
        \caption{Development experience per group}
        \label{fig:dev_experience}
    \end{subfigure}
    ~
    \begin{subfigure}[b]{0.3\textwidth}
        \includegraphics[scale=0.38]{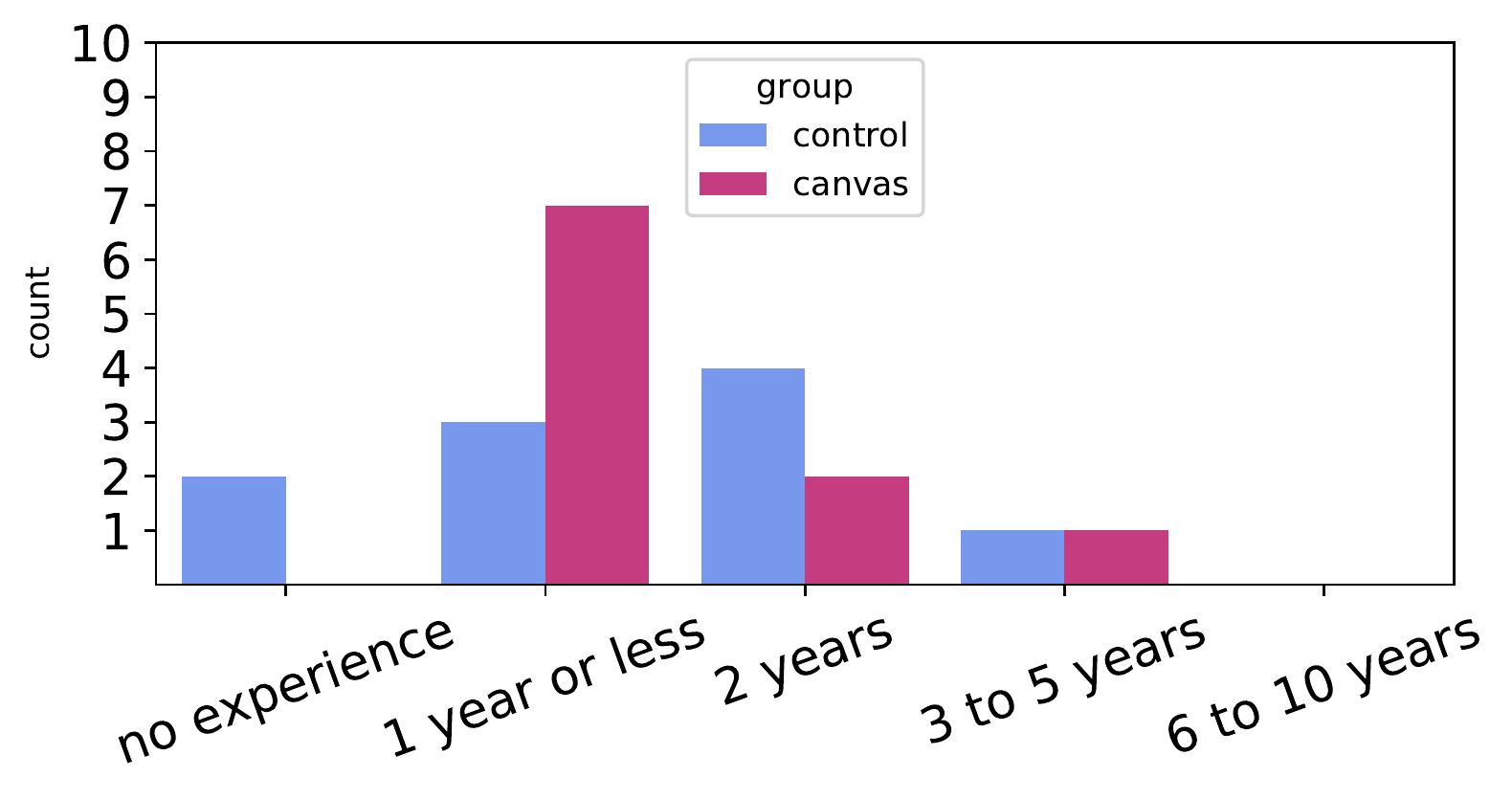}
        \caption{Experience with Dart per group}
        \label{fig:dart_experience}
    \end{subfigure}
    ~
    \begin{subfigure}[b]{0.3\textwidth}
        \includegraphics[scale=0.38]{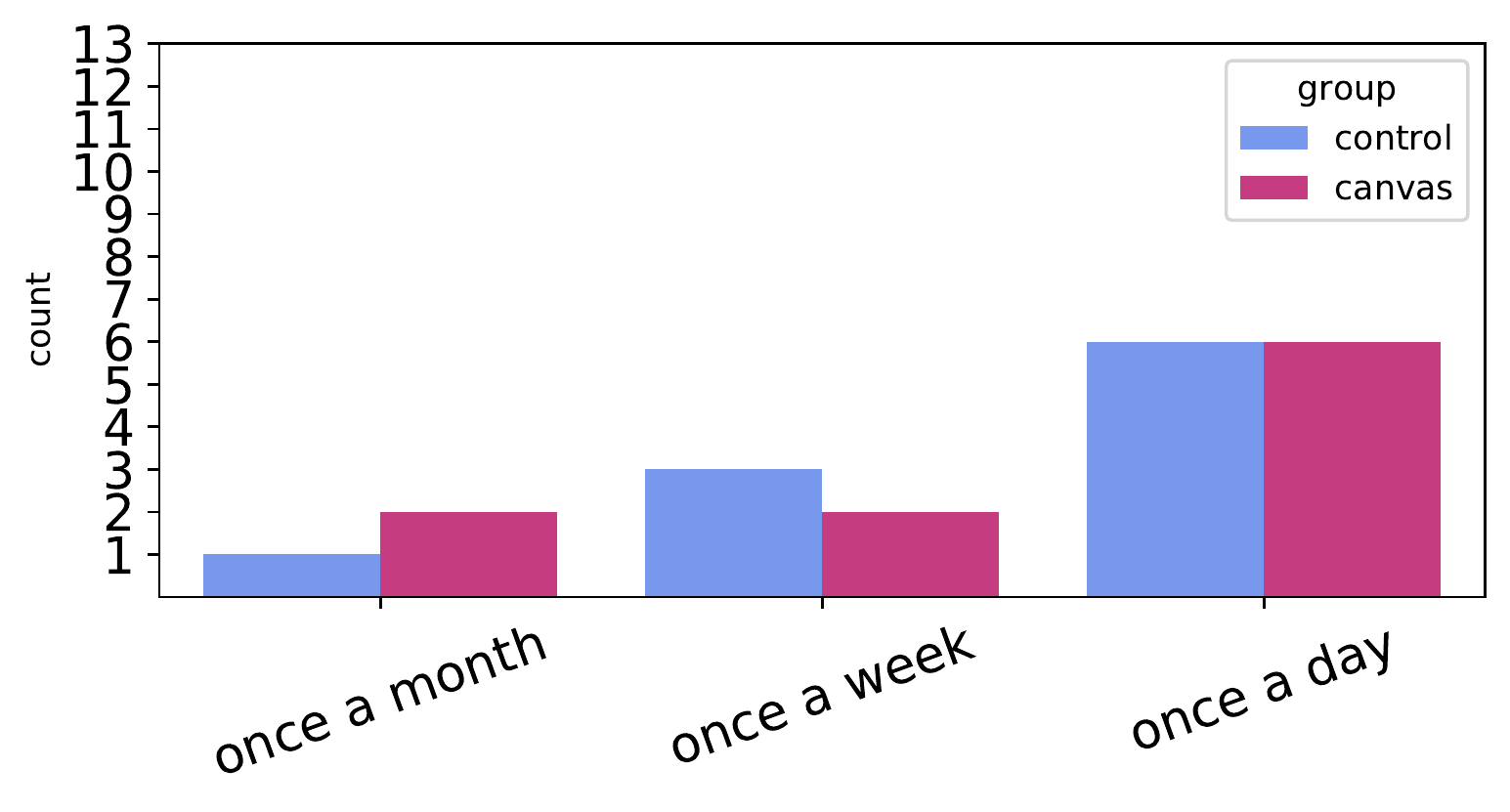}
        \caption{Programming Frequency per group}
        \label{fig:prog_freq}
    \end{subfigure}
    \caption{Experience of the participants in our expriment.}
    \label{fig:experience}
\end{figure*}

\subsection{Procedure}
The participants received an email with the link to the online experiment and their login identifier.
The link gave them access to the first page in which a description of the purpose and steps of the experiment were reiterated and a privacy notice  was outlined.
They were then invited to give their consent by clicking on the ``Next'' button to access the experiment.
Depending on the group they were assigned to, the participants were redirected to a walk-through either of the tab-based version or the canvas version of the tool.
The walk-through consisted of a short description of the main tool features and included GIFs describing the UI elements.

In the next step, participants were introduced to the Flutter Folio code base.
The introduction contained a short description of the main features. 
The participants also had the possibility to test the web version of the app, which was integrated into an iframe.
The introduction concluded with an overview of the app's architecture to help participants navigate the code structure.
Lastly, the participants were given a brief overview of the task structure and phases and then redirected  to the respective version of the tool, where they were prompted to enter their assigned login. 

Having done so, they were presented with a warm-up task designed to familiarise them with the tool. 
Here, they were asked to open a file from the file tree, click on the first occurrence of a given class name, and navigate to its definition using the \textit{Go to Definition} feature.
Inside the class definition file, participants were asked to annotate the first line.
\begin{table*}[]
\caption{Descriptive statistics and results of the Mann-Whitney U test for time and accuracy differences between groups.}
    \centering
    \renewcommand{\arraystretch}{1}
    \begin{tabular}{ll rrr >{\columncolor[gray]{0.9}}r>{\columncolor[gray]{0.9}}r>{\columncolor[gray]{0.9}}r rrr >{\columncolor[gray]{0.9}}r>{\columncolor[gray]{0.9}}r>{\columncolor[gray]{0.9}}r}
    \toprule
    & &\multicolumn{3}{c}{\textbf{Task 1}} & \multicolumn{3}{c}{\textbf{Task 2}} & \multicolumn{3}{c}{\textbf{Task 3}} & \multicolumn{3}{c}{\textbf{Overall Average}} \\
    &	&$T$	&$A_F$&	$A_{AN}$	&$T$	&$A_F$&	$A_{AN}$&$T$	&$A_F$&	$A_{AN}$  &$T$	&$A_F$&	$A_{AN}$\\
    \midrule
    \multirow{2}{*}{\textbf{Control $N=10$}}
    &	$\overline{x}$& 18.96 & 0.57 & 0.48 & 27.59 & 0.53 & 0.41 &  12.43 &   0.88 &  0.80 & 58.97 & 0.63 &               0.51 \\
    &	$s$&  22.30 & 0.17 & 0.13  & 11.65 & 0.19 & 0.24  &  7.68 & 0.11&  0.24 & 32.47 &       0.13 &            0.14 \\
    \midrule
    \multirow{2}{*}{\textbf{Canvas $N=10$}}
    &	$\overline{x}$& 13.80 &  0.60 & 0.49 & 26.07 &  0.36 & 0.30 &  9.08 & 0.81&   0.69 & 48.94 &  0.56 & 0.46\\
    &	$s$& 6.88 & 0.19 &  0.21  &  22.40 & 0.17 &  0.19   &  3.62 & 0.18 &  0.23 & 22.37 & 0.14 & 0.17\\
    \midrule
    \multirow{2}{*}{\textbf{Total $N=20$}}
    &$\overline{x}$ & 16.38 &  0.58   &  0.48 & 26.83  &   0.44  & 0.36  &   10.75  &  0.84   & 0.75  & 53.96 & 0.59 & 0.48\\
    &	$s$ & 16.27 &  0.17 & 0.17  & 17.34 & 0.20  &  0.22 & 6.09  & 0.15 &   0.24  & 27.62 & 0.14 &	0.15\\
\midrule
\textbf{Mann-Whitney U}&&	47 & 42 & 44.5 &  35&	25.5 &	37.5 & 40& 39& 34& 43&	35.5&	40\\
\textbf{Wilcoxon W} && 102 &	97&	99.5&	90 &	80.5&	92.5&	95&	94& 89& 98 & 90.5 & 95\\
\textbf{Z}&&	-.227&	-.615&	-.418&-1.134&-1.865&-.947&-.756&-.880&-1.256&-.529&	-1.100 &	-.757\\
\textbf{\textit{p}}&&	.853&	.579&	.684&	.280&	0.063& .353& .481& .436& .247& .631&	 .280& .481\\
\textbf{\textit{r}}&&0.051& 0.138& 0.093& 0.254& 0.417& 0.212& 0.169& 0.197& 0.281& 0.118& 0.246& 0.169\\
\midrule

\multirow{2}{*}{\parbox{3cm}{\textbf{Kendall's $\tau_b$ Correlation with Corsi Block Score}}}
&\textit{Coeff.}& 	0.083&	0.065&	0.057&	0.028&	.166&	.135&	-.226&	.019& .110 &	-.039&	.079& .162\\
&\textit{p}&  .622&	.713&	.741&	.870&	.337&	.429&	.178&	.917& .538&	.818&	.644& .340\\
\bottomrule
\end{tabular}

\label{tab:descriptive_stats}
\end{table*}

During the experiment, participants were presented with three tasks in sequential order.
In each task, they were asked to make annotations, which appeared in the corresponding task's menu and they could proceed to the next task by clicking on ``continue''.
It was by design not possible to go back to a previous task.
When starting a new task, a new colour for annotations was automatically adopted to better distinguish the annotations.
The files a participant opened during a task remained so for subsequent tasks, i.e., no reset of the user interface was made when starting a new task.
Having completed the tasks, participants were redirected to the Corsi Block test and subsequently answered the demographic survey.

Prior to rolling out the experiment to the actual participants, we tested it with a researcher in our lab.
We gathered feedback about the interaction with the tool and the complexity of the tasks.
Based on this feedback, we adjusted the warm-up task as it was found to be confusing and fixed an issue related to the performance of the \textit{Find all References} feature of the tool.

The results of the experiment, our analysis  scripts, and the details about the online tool are made available online\footnote{https://github.com/abiUni/spatial\_code\_comprehension}.

\section{Results}
\label{sec:results}

\subsection{Demographic Differences}
We start our analysis by looking at the differences in demographic variables between the two groups.
Participants were randomly assigned to the treatment and control groups.
This resulted in a sample of 10 treatment group participants and 10 control group participants.
Figure \ref{fig:experience} showcases the differences in professional development experience, experience with the Dart programming language and programming frequency among the groups.
Participants of the treatment group exhibited a higher mode for  professional development experience (3 to 5 years, n=6) than  the control group (2 years, n=5).
Inversely, the mode for experience with Dart is higher in the control group (2 years, n=4) than in the treatment group (1 year or less, n=7).
Both groups had an equal mode for programming frequency (once a day, n=6) and both groups reported having worked a median of 4 hours before the experiment.
The treatment group showed slightly lower self-reported mental fitness (moderately tired) compared to the control group (neutral) on a scale from 1 (very tired) to 5 (very fit).

\begin{figure*}[h]
    \centering
    \begin{subfigure}[b]{0.32\textwidth}
        \includegraphics[scale=0.5]{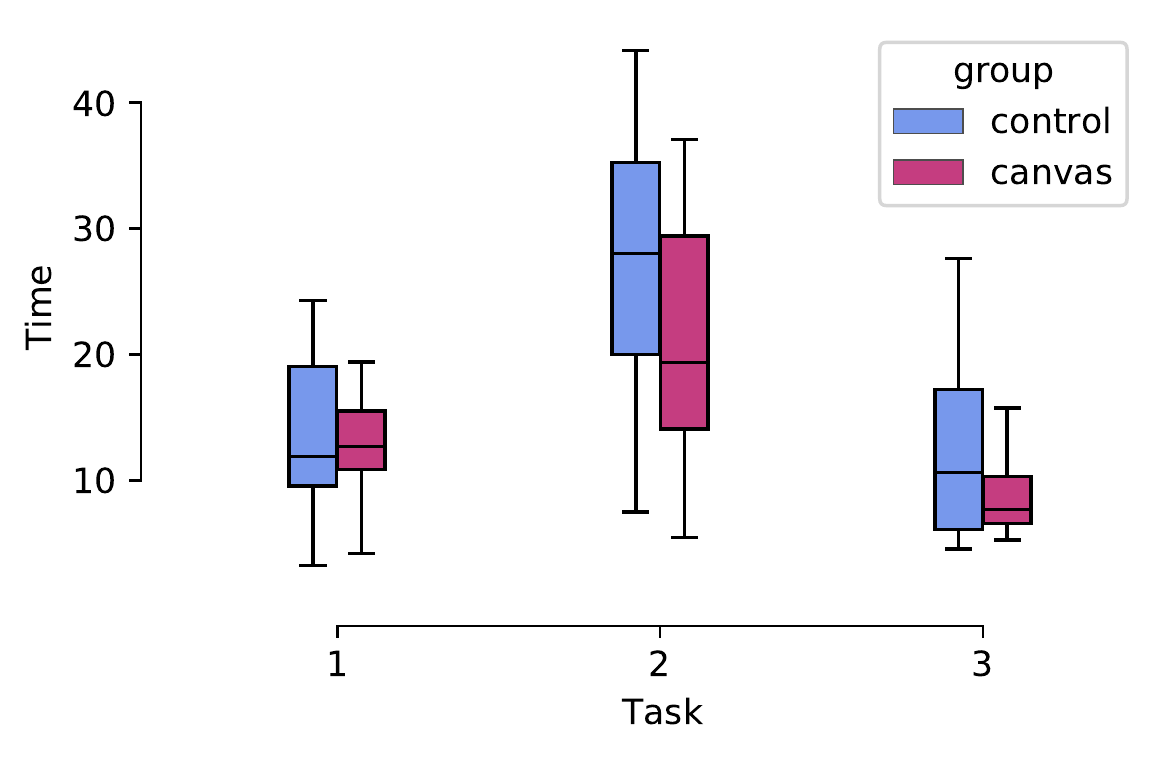}
        \caption{Net time per task}
        \label{fig:boxplot_net_time}
    \end{subfigure}
    ~
    \begin{subfigure}[b]{0.32\textwidth}
        \includegraphics[scale=0.5]{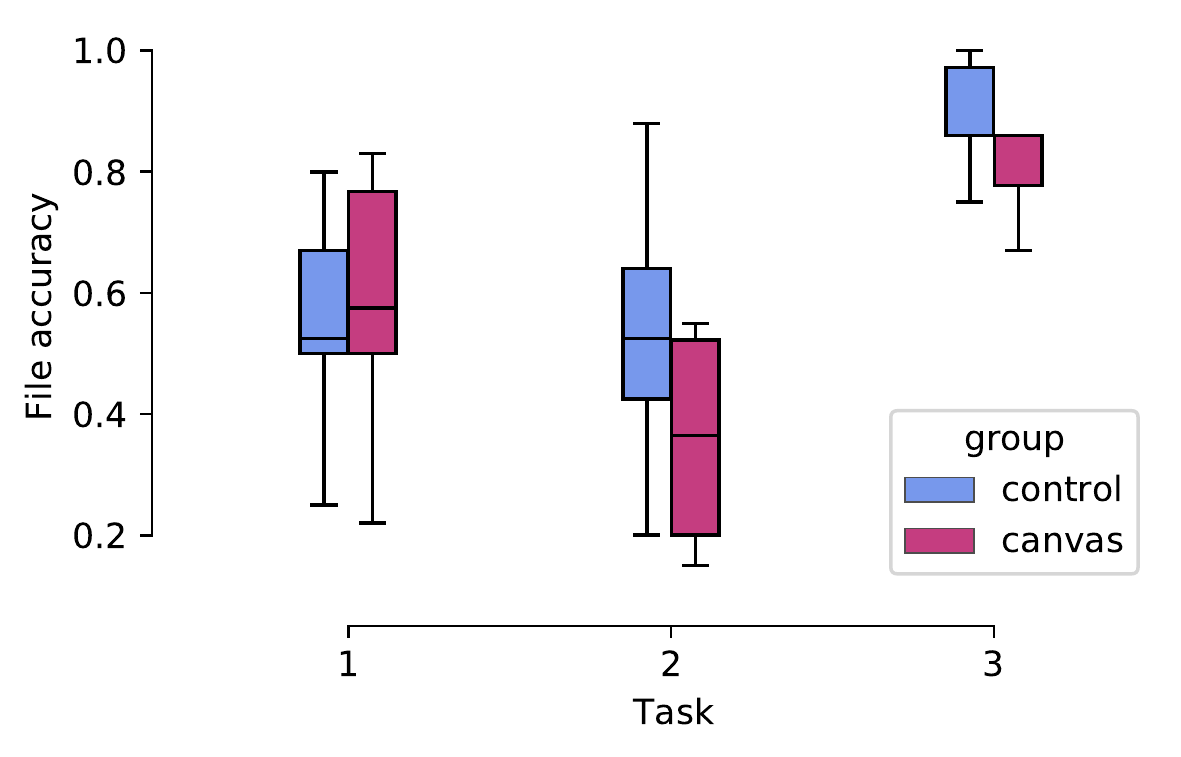}
        \caption{File accuracy per task}
        \label{fig:boxplot_file_accuracy}
    \end{subfigure}
    ~
    \begin{subfigure}[b]{0.32\textwidth}
        \includegraphics[scale=0.5]{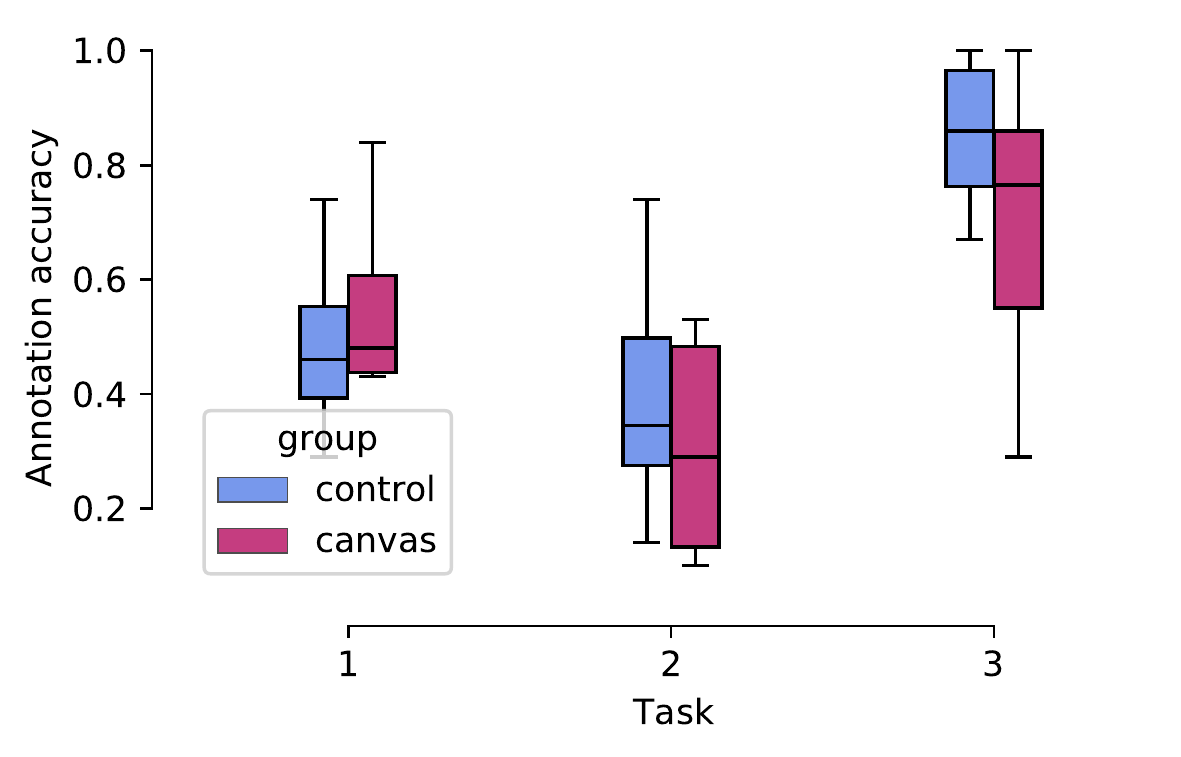}
        \caption{Annotation accuracy per task}
        \label{fig:boxplot_annotation_accuracy}
    \end{subfigure}
    \caption{Boxplots of the comprehension performance variables per experimental task.}
    \label{fig:boxplot_perfromance}
\end{figure*}
\subsection{Spatial Representation of Code (RQ1)}

To answer the first research question, we analyze the difference in performance in terms of time spent on the tasks and the accuracy of the responses between the two groups.
The time is measured as the duration between the point in time when a participant started working on task 1 by clicking the proceed button and the time they clicked on the proceed button after task 3.
We exclude periods of interruptions in which the participants closed the tool window and returned to it at a later point.
The longest interruption time we recorded amounted to 17 hours while all other interruptions lasted less than one minute.
We calculate the accuracy of each participant's response both in terms of relevant files found and correct annotations within these files.
We defined the set of correct annotations as ranges of line numbers for each relevant file.

The number of correct annotations is then calculated as the number of annotated code lines made by the participant that have at least one line of overlap with our solution.
The number of correctly discovered files is calculated as the number of relevant files to which a participant made at least one annotation regardless of whether the correct line range was annotated. 
Following Adeli et al.~\cite{Adeli2020}, we use the Sørensen–Dice F1-score to calculate file and annotation accuracy:

\noindent Accuracy of file selection:
\vspace{-5pt}
\begin{equation}
A_{F} = \frac{2\cdot TP_{F}}{2 \cdot TP_{F} + FP_{F} + FN_{F}}
\end{equation}
\vspace{-5pt}
\noindent Accuracy of annotations:
\vspace{-5pt}
\begin{equation}
A_{AN} = \frac{2\cdot TP_{AN}}{2\cdot TP_{AN} + FP_{AN} + FN_{AN}}
\end{equation}

Where True Positive $TP_{F}$ and $TP_{AN}$ denote the numbers of correctly identified files and annotations respectively.
The False Positive rate $FP_{F}$ for files is the number of annotated files which do not belong to the solution set, while $FP_{AN}$ includes both the number of annotations to irrelevant files and annotations of irrelevant code sections in relevant files.
Lastly, $FN_{F}$ and $FN_{AN}$ is the number of relevant files and annotations not identified by the participant.

Table \ref{tab:descriptive_stats} outlines the mean and standard deviation of the variables in question for the two groups per task as well as cumulatively.
On average, the control group spent more time on each task than the canvas group, with the highest discrepancy in task 1 amounting to 5.16 min.
Both groups have comparable file accuracy except on task 2 where the control group scored 0.53 compared to 0.36 for the canvas group.
We should note that task 2 has the highest number of solution files to be found (7 files).
The control group also had a higher annotation accuracy for tasks 2 and 3 ($A_N = 0.41$ and $A_N = 0.8$ respectively) compared to marginally higher accuracy in the canvas group for task 1 ($A_N = 0.49$).
As depicted in Figure \ref{fig:boxplot_perfromance}, the control group shows a wider spread of the results among its participants, especially for task 2 compared to the canvas group.
Overall, the canvas group spent less time on average on the tasks (48.94 min compared to 58.97 min for the control group) while the control group  showed a slightly higher file accuracy ($A_F = 0.63$) and higher annotation accuracy ($A_N = 0.51$).

To further investigate the significance of these differences, we conducted a Mann-Whitney U test, given that performance variables were not consistently normally distributed in each group.
The null hypothesis of this test is that both samples have an identical distribution.
The results are shown in Table \ref{tab:descriptive_stats}. 
The test does not deliver a significant difference between the groups for the performance variables for any of the tasks ($p>0.05$).
We conclude that the use of a code canvas has no significant effect on comprehension performance despite allowing participants to investigate multiple files simultaneously.
To further investigate this finding, we decided to take a closer look at the control group participants who used the split screen view feature in the tab-based version of the tool.
Only five out of ten participants of this group made use of this feature throughout the experiment, mainly to answer task 3 where it was required to differentiate two API calls.
However, we did not find significant differences in performance variables between the participants of the control group who used the split screen view and those who did not.

\begin{figure*}
    \centering
    \includegraphics[scale=0.37]{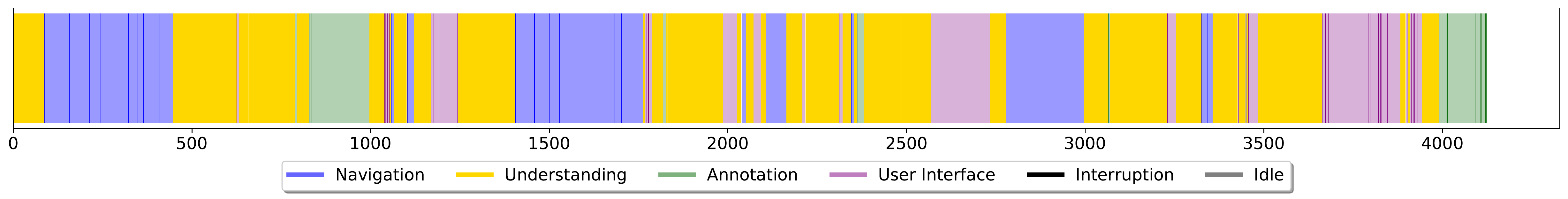}  \caption{Example of time distribution by development activity of one participant in the control group.}
    \label{fig:timeline_activity}
\end{figure*}

\subsection{Comprehension Activities (RQ2)} 

To evaluate the differences in the time allocated to different activities between the two groups, we took inspiration from Minelli et al.'s approach for analyzing interaction data \cite{Minelli2015InteractionData}.
Our analysis focuses on the logged events of user interactions, which we categorised into \textit{Navigation}, \textit{Annotation}, and \textit{UI} activities.
Navigation events include opening a file from the file tree or by using the \textit{Go to Definition} or \textit{Find all References} functionality.
Annotation activities include making an annotation to a code section or removing an annotation.
Lastly, UI activities include moving a file node on the canvas in the canvas version, switching between tabs in the control version  or hiding a file in both tool versions.

\begin{table}[t]
\caption{Time components per participant in minutes.}
    \centering
    \begin{tabular}{llrr rrrr }
\toprule
& \textbf{Participant} &  \textbf{Underst.} &  \textbf{Annotation}  & \textbf{Navigation} & \textbf{UI} \\
\midrule


\multirow{11}{*}{\rotatebox[origin=c]{90}{\textbf{Canvas}}}
&P1&15.51 &     8.46 &       12.42 &            2.12  \\
&P2&17.65 &     0.12 &       19.66 &            0.72  \\
&P3&13.90 &     3.32 &        9.88 &            2.38  \\
&P4&32.61 &    14.99 &        3.66 &            4.08  \\
&P5&16.47 &     2.35 &        3.60 &            0.62 \\
&P6&29.15 &     2.81 &       15.71 &            5.23  \\
&P7&22.45 &    10.21 &        4.93 &            3.44  \\
&P8&31.52 &     7.32 &       11.05 &            2.89  \\
&P9&20.46 &     2.87 &       17.55 &            1.07 \\
&P10&12.57 &     2.75 &        3.17 &            0.55  \\

& \textbf{Mean}&  \textbf{21.23}& 	\textbf{5.52}& 	\textbf{10.16}& 	\textbf{2.31} \\
\midrule
\multirow{11}{*}{\rotatebox[origin=c]{90}{\textbf{Control}}}
&P11&17.20 &     4.18 &       11.42 &            0.64 \\
&P12&36.21 &     8.06 &        5.84 &            1.47  \\
&P13&56.17 &     4.79 &       42.32 &            3.10\\
&P14&33.74 &     5.70 &       17.91 &           11.36  \\
&P15&19.03 &     2.79 &       29.89 &            0.58  \\
&P16&12.34 &     1.25 &        5.40 &            4.17  \\
&P17&32.19 &     0.00 &       14.02 &            6.83  \\
&P18&22.36 &     4.14 &        9.25 &            2.05  \\
&P19&16.82 &     0.73 &       13.20 &            1.58  \\
&P20&16.92 &     0.81 &       12.02 &            0.00 \\

& \textbf{Mean}&  \textbf{26.30}& 	\textbf{3.25}& 	\textbf{16.13}& 	\textbf{3.17} \\
\midrule
\multicolumn{6}{c}{\textbf{Mann-Whitney U Test}}\\
\midrule
\multicolumn{2}{l}{\textbf{{Mann-Whitney U}}}&38&37&33&49\\
\multicolumn{2}{l}{\textbf{{Wilcoxon W}}}&93&92&88&104\\
\textbf{\textit{Z}}&&-.907&-.983&-1.285&-.076\\
\textit{\textbf{p}}&&.393&.353&.218&.971\\
\textbf{\textit{r}}&&0.2& 0.22& 0.29& 0.02\\
\midrule
\multicolumn{6}{c}{\textbf{2-tailed Kendall's $\tau_b$ Correlation with Corsi Block Score}}\\
\midrule
\multicolumn{2}{l}{\textit{Coeff.}}&-.083&	-.458&	.314&	-.436\\
\multicolumn{2}{l}{\textit{p}}&.622&	\textbf{.006**}&	.061&	\textbf{.009**}\\
\midrule
\multicolumn{6}{c}{\textbf{1-tailed Kendall's $\tau_b$ Correlation with Corsi Block Score}}\\
\midrule
\multicolumn{2}{l}{\textit{{Coeff.}}} &  -.083 & -.458 & .314 & -.436\\
\multicolumn{2}{l}{\textit{{p}}} & .311&	\textbf{0.003**}&	\textbf{0.031*}&	\textbf{0.005**}\\
\bottomrule
\end{tabular}
    
    \label{tab:time_component}
\end{table}

In their work, Minelli et al.~\cite{Minelli2015InteractionData} used granular interactions such as mouse clicks, mouse drifts and keyboard strokes to detect developer actions after a trigger event for a certain activity occurs.
For example, opening a file from the file tree triggers the start of a navigation activity.
The authors then aggregated these fine-grained interactions showing a time distance less than one second apart. 
This duration refers to the reaction time (RT) also commonly known as the Psychological Refractory Period \cite{RT52} representing the response delay made by humans when asked to react to two consecutive stimuli with decreasing interleaving time.

The applicability of this approach in our experiment is limited by the fact that we do not track fine-grained interactions from the mouse and keyboard.
However, we follow a similar approach to estimating the time dedicated to each activity by  aggregating trigger events belonging to the same development activity. 
A visualization of these aggregates is provided in Figure \ref{fig:timeline_activity}.
We differentiate between interruptions occurring when a participant closes the web tool and logs in again and idle time during which the participant does not produce any event.
We assume a minimum idle time of 10 minutes inspired by Minelli et al.  \cite{Minelli2014}.
We determine the total duration of the \textit{Navigation}, \textit{Annotation} and \textit{UI} activities by summing up the duration between the corresponding events that are less than the minimum idle time and which are not separated by an interruption period.
As defined by Minelli et al. \cite{Minelli2015InteractionData}, the basic inter-activity understanding time is the sum of time intervals between subsequent activities longer than one RT and shorter than the minimum idle time.

Table \ref{tab:time_component} outlines the  time spent by each participant on the development activities.
The total times are calculated as the duration from the first  until the last event, from which we subtract interruptions and idle times.
On average, the canvas group spent less time on understanding  and navigation than the control group.
The largest difference between the two groups was observed for navigation ($\Delta = 5.97 $ min) followed by understanding ($\Delta = 5.07 $ minutes).
The control group on the other hand spent more time interacting with the UI and less time making code annotations.
The time differences for the latter two activities are less pronounced between the groups ($\Delta = 2.27$ min for annotation and $\Delta = 0.86$ min for UI interactions).
These differences did not prove significant ($p>0.05$) for any of the development activities as reported by the Mann-Whitney U test in Table \ref{tab:time_component}.

\subsection{Visuo-spatial Working Memory (RQ3)}
\begin{figure}[b]
    \begin{subfigure}[b]{0.2\textwidth}
        \centering
        \includegraphics[scale=0.35]{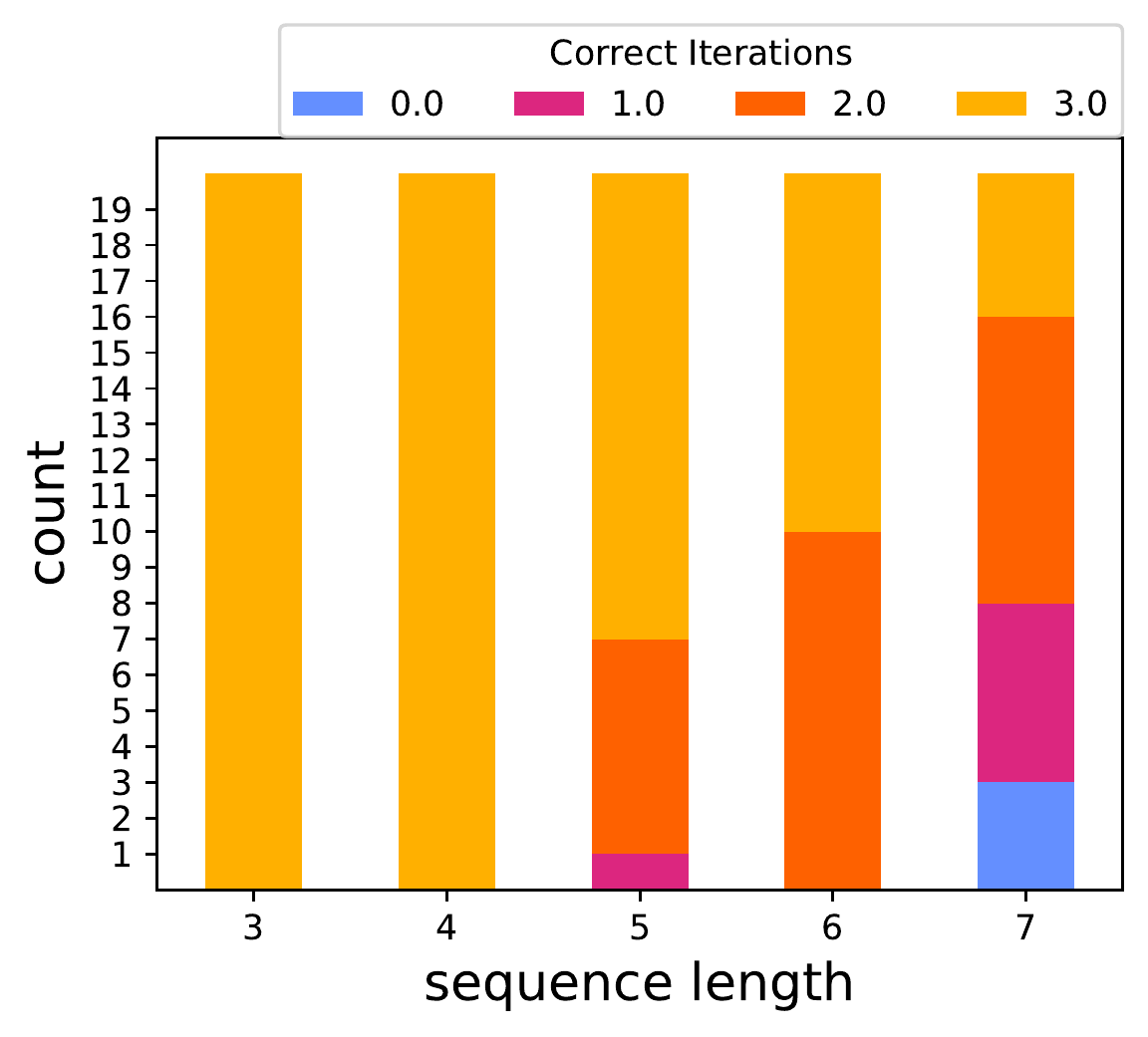}
    \end{subfigure}
    ~
    \begin{subfigure}[b]{0.11\textwidth}
        \centering
        \includegraphics[scale=0.5]{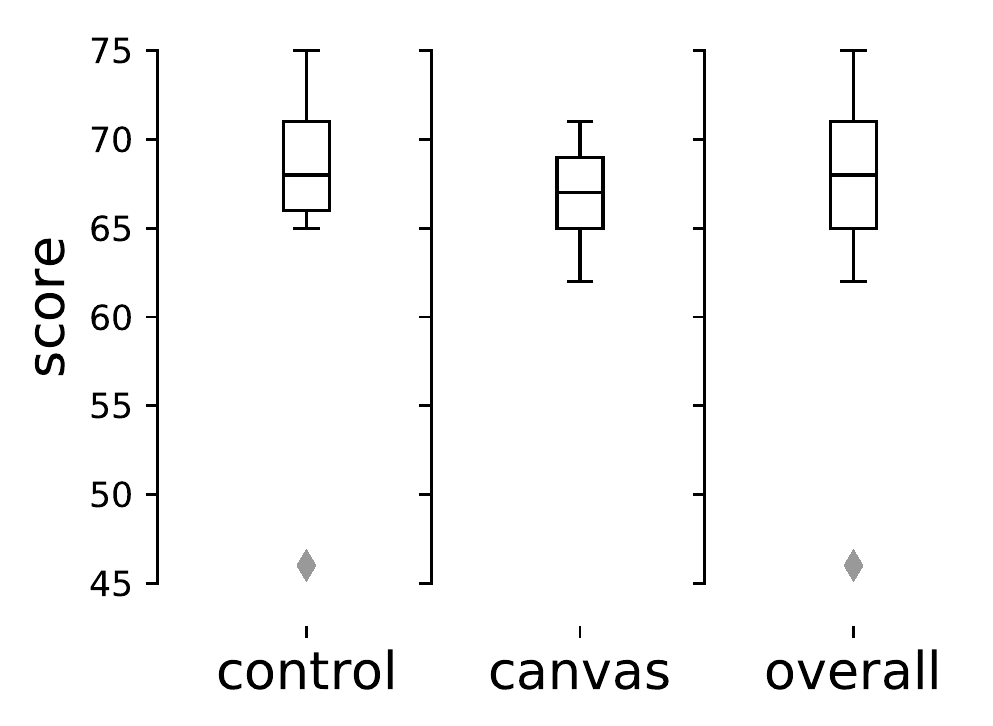}
    \end{subfigure}

    \caption{Results of the Corsi Block tests.}
    \label{fig:corsi_results}
    
\end{figure}

To answer the third research question, we analyzed the correlation between each participant's score on the Corsi Block test and their performance on the comprehension tasks.
The mean score of the participants in our sample was 67 with a standard deviation of 6.05.
The box plots shown in Figure \ref{fig:corsi_results} outline the differences in Corsi Block test scores between the two groups.
The median score for the control group (68) is slightly higher than in the canvas group (67).
Only one outlier was registered for a participant of the control group who scored 46 points.
The number of correct iterations recalled by participants diminishes with increasing sequence length as can be seen in Figure \ref{fig:corsi_results}.
All participants were able to correctly answer all three iterations of sequence lengths 3 and 4.
For sequence lengths of 5 and 6, this number dropped to 13 and 10 participants respectively.
Only 4 participants were able to recall all three sequences of length 7 and 3 participants failed to recall any of these sequences correctly.

We then took a  look at the correlation between a participant's score on the Corsi Block test and their performance on the code comprehension tasks.
We used a 2-tailed Kendall’s $\tau_{B}$ correlation coefficient.
Our null hypothesis is that no correlation exists between a participant's visuo-spatial working memory and their performance reflected by the accuracy of their answers and the time they spent on the tasks.
The results of the correlation analysis are shown at the end of Table \ref{tab:descriptive_stats}.
The highest magnitude of the correlation coefficient is registered for the time variable on task 3 ($c=-.226, p=0.187$).
However, this correlation remains insignificant using a 1-tailed Kendall’s $\tau_{B}$ ($p=0.089$).
Looking at the cumulative variables over all tasks, we can see that the coefficient is the highest for annotation accuracy (0.162) and comparatively low for both file accuracy and time variables (0.079 and -0.039).
These correlations are, however, also not significant at the 95\% confidence interval ($p>0.05$).
We, therefore, fail to reject the null hypothesis stating no correlation between performance variables and the Corsi Block test score.

We also investigate the correlation between the time spent on each comprehension activity and the Corsi Block test score.
Table  \ref{tab:time_component} shows a significant negative correlation between the Corsi Block test score and the time spent on  annotation and UI interactions at the 0.01 level.
Only time spent on navigation correlated positively with the visuo-spatial working memory score.
A 1-tailed Kendall’s $\tau_{B}$  (Table \ref{tab:time_component}) test furthermore  showed a significant correlation with navigation time ($ c=0.314, p=0.031$) indicating a medium to  strong association of the two variables.
The results thus suggest that participants with a higher visuo-spatial memory made fewer annotations, interacted less with the UI and navigated to more files.

\begin{figure*}[!h]
    \centering
    \includegraphics[scale=0.44]{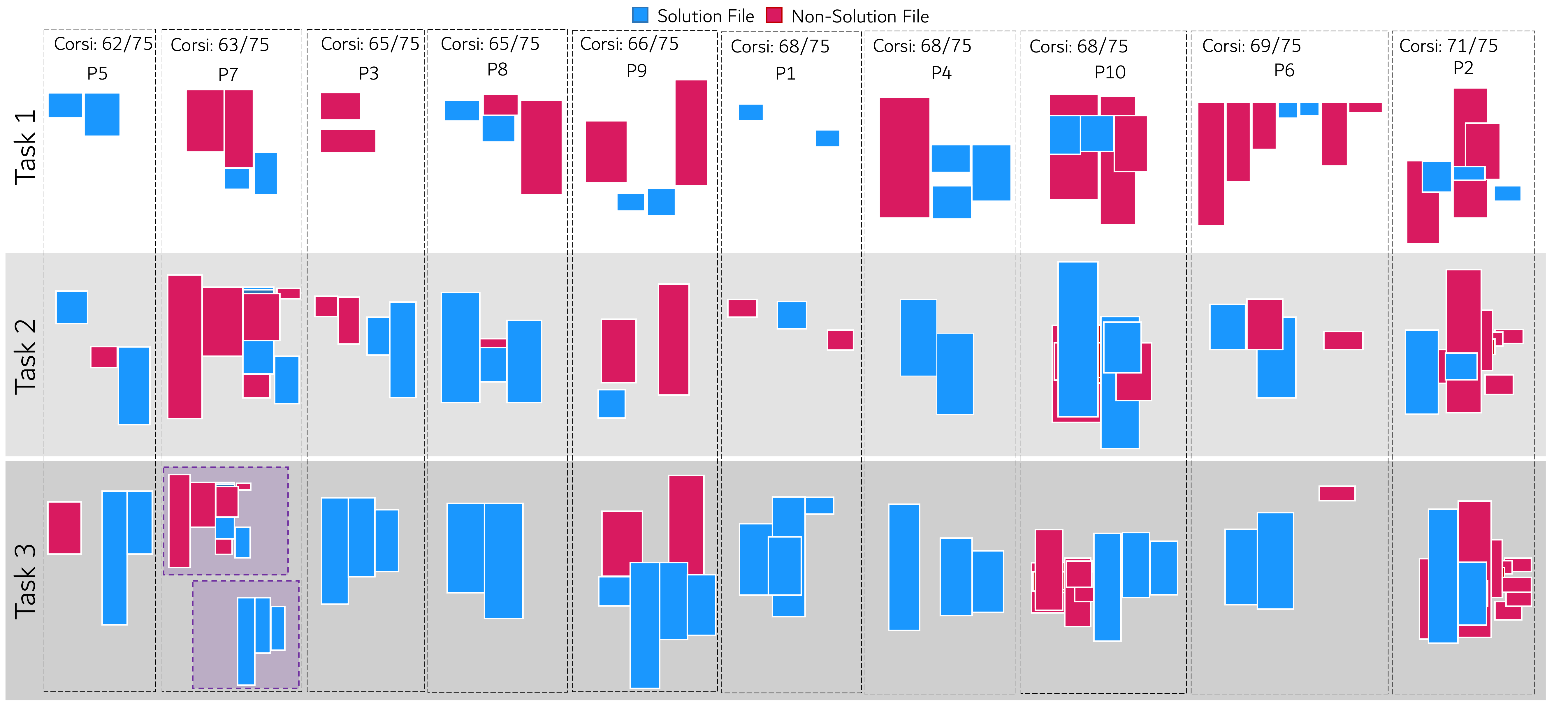}
    \caption{Visual layouts of participants' code canvases at the end of each task. Blue files are relevant for solution, red are not.}
    \label{fig:canvas_label}
\end{figure*}

Finally, we also qualitatively analyze the  usage of the code canvas in the treatment group.
We used the activity timeline and the file positions to reconstruct the state of the canvas and the distribution of the files at the end of each task.
Figure \ref{fig:canvas_label} illustrates the thus constructed canvas state. 
The boxes representing the files were scaled to fit available space and do not convey the original dimensions.
Two clusters emerge from the canvas states as evidence that participants of this group used the canvas differently.
In fact, we can see that a subgroup of participants kept only a few files open simultaneously (this includes participants $P1, P3, P4, P5$)  while another subgroup had visibly more files open at the end of each task ($P2, P6, P7, P8, P9, P10$).

By ordering the participants according to an ascending Corsi Block test score, we can see that participants with lower scores (62 to 68 out of 75) tend to ``clean up their canvas'' by keeping fewer files open  and in some instances ordering them in a way that avoids overlaps.
This is  the case for $P8$ in task 1 and $P9$ and $P5$ for tasks 1 and 2.
Another strategy followed by $P7$ consisted in moving the focus to an empty part of the zoomable canvas to start a new task (in this case task 3).
On the other end of the spectrum, we noticed a relatively higher number of files, which did not seem to have been arranged in any particular order, evidenced by the overlaps.
This is the case, for example, for $P2$ and $P10$, in tasks 1 and 2.
This observation  might help explain the positive correlation between visuo-spatial working memory and time spent on navigation as participants who scored higher on the Corsi Block navigated to more files and spent less time on UI interactions to arrange the files.

\subsection{Effect of Measured Extraneous Variables}
Extraneous variables have no significant correlation with the results
with the exception of a negative correlation between professional development experience  and overall time spent on the tasks ($c=-0.396, p=0.03$).
However, the linear regression models of the time variable controlling for the effects of extraneous variables have poor explanation power ($R^2< 0.5$) as described in Table \ref{tab:regression_model}.
In particular, professional development experience did not contribute a large increase in the value of $R^2$ (from $0.035$ to $0.101$) in these regression models.
\begin{table}[h]
    \centering
\caption{Regression models for time spent on the tasks.}
    \renewcommand{\arraystretch}{1.3}
    \begin{tabular}{lrr}
    
    \toprule
    \textbf{Predictor} & \textbf{$R^2$} & \textbf{Sig.}\\
    \midrule
    Group & 0.035 & .432\\
    + Professional Development experience & .101& .406\\
    + Dart Experience & .180& .353\\
    + Programming Frequency & .186&.512\\
    + Work Hours & .186&.672\\
    + Mental Fitness & .197&.776\\
    + Corsi Block score & .213&.842\\
    \bottomrule
    \end{tabular}

    \label{tab:regression_model}
\end{table}

\section{Discussion of Findings and their Implications}
\label{sec:discussion}

Our results suggest that developers' visuo-spatial mental model does not associate significantly with 
comprehension performance but that developers' behaviour in development environments may differ based on this factor. 
The treatment group in our experiment needed on average less time to complete code comprehension tasks ($\Delta \approx 10$ min).
However, the differences in file and annotation accuracy are not significant.
Our analysis of time spent on different development activities has revealed that control group participants spent more time on navigation, understanding, and UI interactions, than the canvas group.

Despite being statistically insignificant, we can partly attribute these differences to the control group participants having a limited number of files in their window focus compared to the canvas group.
In fact, even using the split screen view, participants of this group opened a maximum of 2 files simultaneously.
Thus, changing the focus to another file requires either switching to the corresponding tab or re-opening the file from the file tree.
Additionally, the tab-based version of the tool requires the user to scroll within each of the split-screen windows to inspect the  file contributing to higher understanding time.
A code canvas partly alleviates this limitation but still requires the user to zoom in and out to switch between the focus on a single file and the overview of multiple files.
When designing future code navigation tools, a hybrid visualization could allow the user to enter a full-focus mode for single files on a canvas.

The reconstruction of the canvas timeline shown in Figure \ref{fig:canvas_label} allowed us to see that some of the canvas group participants took care of arranging the files on the canvas in a way that suggests they were looking at logical relatedness.
This is particularly apparent for $P8$ and $P4$ in task 1.
We also came across multiple instances (for example $P4$, $P6$) where the two files to be compared in task 3,  \textit{NativeFirebaseService} and \textit{DartFirebaseService}, were aligned side by side in a way that made it easier to compare the code line responsible for the API call.
This suggests that a code canvas can  allow a developer to study the dependencies between multiple candidate files at a time.
Moreover, by preserving the spatial arrangement of the files defined by the user, it can contribute to the creation of cues to help memorise related code sections.

We found no significant correlation between visuo-spatial working memory and code comprehension performance. 
However, a positive correlation between visuo-spatial working memory and time spent on navigation corroborates with the observations of Jones and Burnett~\cite{Jones2007SpatialSkills}.
The authors studied student performance on code comprehension tasks in relation to the score on a mental rotation test.
They reported that students with higher spatial skills carried out relatively more jumps between class files, which can be considered part of navigation activities.
Moreover, a significant negative correlation of the Corsi Block test score with time spent on UI interactions and annotations suggests that people with higher visuo-spatial working memory tend to navigate through code more efficiently by making less tab-switches and more targeted annotations.

The canvas layouts also show that individuals with relatively high score on the Corsi Block test had more files open on the canvas at the end of each task.
The files did not appear to have been arranged in any specific order, seemingly resulting in visual clutter most visible for $P2$ and $P10$ for tasks 2 and 3.
A study by Minelli et al.~\cite{Minelli16TamingIDE} analyzing the number of open IDE windows in development sessions showed that it is commonplace for developers to work in a chaotic environment.
By their estimation, developers spend on average more than 30\% of their time in an overloaded environment with high overlap between windows requiring more than 100\% of the available screen size.
This problem is thus not specific to the code canvas but nevertheless affects its usage.
In this case, the decision is left to the developers whether or not to invest time in arranging the files, which they may want to recall after the end of a comprehension session.
As suggested by our results, this might prove more beneficial for individuals with lower visuo-spatial memory to help build visual cues.

In general, our results suggest that cognitive aspects such as visuo-spatial working memory should be considered 1) a potential confounding factor when designing future comprehension studies as well as 2) when designing personalised tool support for developers since such cognitive aspects seem to affect developers' behavior and their allocation of effort \cite{Penta:ICPC:2007, Lucca:ICPC:2006, Wagner:TSE:2021}.

\section{Threats to Validity}
\label{sec:threats}
\textbf{External Validity}
The small sample size in our experiment limits the generalisability of our results and may be the cause of statistical insignificance in the results of  RQ1 and RQ2.
However, the participants in our sample have a diverse backgrounds and either worked as professional Flutter developers or were students working part-time in industry.

\vspace{6pt}
\textbf{Internal Validity}
We resorted to randomization instead of stratification as well as a between-subject design instead of a within-subject design to avoid further reducing the number of participants per group, which would further weaken statistical power.
Moreover, our analysis of extraneous variables between the groups showed that while differences exist, they do not significantly influence the results.
Unlike previous studies, \cite{Minelli2015InteractionData}, we did not log fine-grained user interactions such as mouse drifts and keyboard strokes.
As a result, the estimated idle time may have been exaggerated leading to an underestimation of understanding time, particularly for the canvas group as zoom-in and out events were not logged.
However, we minimised this risk by choosing a threshold of 10 min, which is  close to the estimated effort to complete a single task.
Furthermore, having the Corsi Block test at the end of the experiment meant that we are unable to  detect measurement errors caused by fatigue. 
However, given that the mean test scores were relatively high across both groups, this does not seem to be a major problem.
Furthermore, participants' file arrangements might have been influenced by their knowledge that they will not be using the canvas tool in the future and thus did not see the need to invest time in ordering the files in a meaningful way.
The chosen tasks may not be representative of development activities.
To mitigate this threat, we included two phases in each task with the first one focusing mainly on navigation.
Moreover, task 2 was taken from a resolved issue in the Flutter Folio app.
Lastly, since we used a web tool for the experiment, latencies related to connection quality might have been experienced differently between participants.

\vspace{6pt}
\textbf{Construct Validity} The Corsi block test might not have been an adequate test of spatial ability on its own, as it is sometimes used in combination with other tests such as mental rotation. 
This would, however, have made the experiment last longer. 
The Corsi block test assesses the ability to memorise the location of elements and to navigate an environment. 
In our setting, the environment in question is the unfamiliar code base and the comprehension tasks require participants to engage in information foraging and inevitably memorise the location of certain functionalities.

\section{Conclusion}
\label{sec:conclusion}
An in-depth grasp of  program comprehension and the factors that affect it is one of the major research challenges of the Software Engineering community.  
Our work specifically focuses on the relationship between spatial code representation, visuo-spatial working memory, and code comprehension. 
We found no significant difference in comprehension performance and in time spent on different development activities between using tab-based and canvas-based representations. 
However, we found a significant negative correlation between developers' visuo-spatial memory and time spent on UI interactions and annotating as well as a positive correlation with the time spent on navigation.

Our observations also suggest an association between code navigation behavior and visuo-spatial working memory -- evidenced by the number of files opened and the layout of the resulting canvas.
Our observations corroborate with previous research mainly investigating window plague related to inefficient navigation in IDEs.
Further research should give more attention to cognitive characteristics as confounding variables in the design of code comprehension studies and to the personalisation of navigation  support in the IDE.

\section*{Acknowledgment}
We would like to thank all study participants. 
This work was funded by the Deutsche Forschungsgemeinschaft (DFG, German Research Foundation) - Project Number: 166725071.

\IEEEtriggeratref{38}
\bibliographystyle{IEEEtran}

\begin{thebibliography}{10}
\providecommand{\url}[1]{#1}
\csname url@samestyle\endcsname
\providecommand{\newblock}{\relax}
\providecommand{\bibinfo}[2]{#2}
\providecommand{\BIBentrySTDinterwordspacing}{\spaceskip=0pt\relax}
\providecommand{\BIBentryALTinterwordstretchfactor}{4}
\providecommand{\BIBentryALTinterwordspacing}{\spaceskip=\fontdimen2\font plus
\BIBentryALTinterwordstretchfactor\fontdimen3\font minus
  \fontdimen4\font\relax}
\providecommand{\BIBforeignlanguage}[2]{{%
\expandafter\ifx\csname l@#1\endcsname\relax
\typeout{** WARNING: IEEEtran.bst: No hyphenation pattern has been}%
\typeout{** loaded for the language `#1'. Using the pattern for}%
\typeout{** the default language instead.}%
\else
\language=\csname l@#1\endcsname
\fi
#2}}
\providecommand{\BIBdecl}{\relax}
\BIBdecl

\bibitem{Minelli2014}
R.~Minelli, A.~Mocci, M.~Lanza, and L.~Baracchi, ``Visualizing developer
  interactions,'' in \emph{2014 Second IEEE Working Conference on Software
  Visualization}, 2014, pp. 147--156.

\bibitem{Maalej:TOSEM:14}
\BIBentryALTinterwordspacing
W.~Maalej, R.~Tiarks, T.~Roehm, and R.~Koschke, ``On the comprehension of
  program comprehension,'' \emph{ACM Transactions on Software Engineering and
  Methodology}, vol.~23, no.~4, sep 2014. [Online]. Available:
  \url{https://doi.org/10.1145/2622669}
\BIBentrySTDinterwordspacing

\bibitem{SiegmundCC2016}
J.~Siegmund, ``Program comprehension: Past, present, and future,'' in
  \emph{2016 IEEE 23rd International Conference on Software Analysis,
  Evolution, and Reengineering (SANER)}, vol.~5, 2016, pp. 13--20.

\bibitem{siegmund2017measuring}
J.~Siegmund, N.~Peitek, C.~Parnin, S.~Apel, J.~Hofmeister, C.~K{\"a}stner,
  A.~Begel, A.~Bethmann, and A.~Brechmann, ``Measuring neural efficiency of
  program comprehension,'' in \emph{Proceedings of the 2017 11th Joint Meeting
  on Foundations of Software Engineering}, 2017, pp. 140--150.

\bibitem{peitek2021program}
N.~Peitek, S.~Apel, C.~Parnin, A.~Brechmann, and J.~Siegmund, ``Program
  comprehension and code complexity metrics: An fmri study,'' in \emph{2021
  IEEE/ACM 43rd International Conference on Software Engineering (ICSE)}.\hskip
  1em plus 0.5em minus 0.4em\relax IEEE, 2021, pp. 524--536.

\bibitem{hofmeister2017shorter}
J.~Hofmeister, J.~Siegmund, and D.~V. Holt, ``Shorter identifier names take
  longer to comprehend,'' in \emph{2017 IEEE 24th International conference on
  software analysis, evolution and reengineering (SANER)}.\hskip 1em plus 0.5em
  minus 0.4em\relax IEEE, 2017, pp. 217--227.

\bibitem{Peitek:FSE:2022}
``Correlates of programmer efficacy and their link to experience: A combined
  eeg and eye-tracking study,'' in \emph{Proceedings of the ACM Joint European
  Software Engineering Conference and Symposium on the Foundations of Software
  Engineering (ESEC/FSE)}.\hskip 1em plus 0.5em minus 0.4em\relax ACM, November
  2022.

\bibitem{kersten2006using}
M.~Kersten and G.~C. Murphy, ``Using task context to improve programmer
  productivity,'' in \emph{Proceedings of the 14th ACM SIGSOFT international
  symposium on Foundations of software engineering}, 2006, pp. 1--11.

\bibitem{DeLineCodeCanvas2010}
R.~DeLine and K.~Rowan, ``Code canvas: zooming towards better development
  environments,'' in \emph{2010 ACM/IEEE 32nd International Conference on
  Software Engineering}, vol.~2, 2010, pp. 207--210.

\bibitem{BragdonCodeBubbles2010}
A.~Bragdon, S.~P. Reiss, R.~Zeleznik, S.~Karumuri, W.~Cheung, J.~Kaplan,
  C.~Coleman, F.~Adeputra, and J.~J. LaViola, ``Code bubbles: rethinking the
  user interface paradigm of integrated development environments,'' in
  \emph{2010 ACM/IEEE 32nd International Conference on Software Engineering},
  vol.~1, 2010, pp. 455--464.

\bibitem{DeLineCollab2012}
R.~DeLine, A.~Bragdon, K.~Rowan, J.~Jacobsen, and S.~P. Reiss, ``Debugger
  canvas: Industrial experience with the code bubbles paradigm,'' in \emph{2012
  34th International Conference on Software Engineering (ICSE)}, 2012, pp.
  1064--1073.

\bibitem{Patch2014}
\BIBentryALTinterwordspacing
A.~Z. Henley and S.~D. Fleming, ``The patchworks code editor: Toward faster
  navigation with less code arranging and fewer navigation mistakes,'' in
  \emph{Proceedings of the SIGCHI Conference on Human Factors in Computing
  Systems}, ser. CHI '14.\hskip 1em plus 0.5em minus 0.4em\relax New York, NY,
  USA: Association for Computing Machinery, 2014, p. 2511–2520. [Online].
  Available: \url{https://doi.org/10.1145/2556288.2557073}
\BIBentrySTDinterwordspacing

\bibitem{Siegmund2015Confounders}
J.~Siegmund and J.~Schumann, ``\BIBforeignlanguage{en}{Confounding parameters
  on program comprehension: a literature survey},''
  \emph{\BIBforeignlanguage{en}{Empir. Softw. Eng.}}, vol.~20, no.~4, pp.
  1159--1192, Aug. 2015.

\bibitem{Bancifra22VisuospatialMemory}
J.~J.~Bancifra, {Tarlac State University, Tarlac City, Philippines}, and
  {https://orcid.org/0000-0003-0641-1305}, ``Supervisory practices of
  department heads and teachers' performance: Towards a proposed enhancement
  program,'' \emph{APJAET - Journal ay Asia Pacific Journal of Advanced
  Education and Technology}, pp. 25--33, Sep. 2022.

\bibitem{DeLineThumbnails2006}
R.~DeLine, M.~Czerwinski, B.~Meyers, G.~Venolia, S.~Drucker, and G.~Robertson,
  ``Code thumbnails: Using spatial memory to navigate source code,'' in
  \emph{Visual Languages and Human-Centric Computing (VL/HCC'06)}, 2006, pp.
  11--18.

\bibitem{Adeli2020}
M.~Adeli, N.~Nelson, S.~Chattopadhyay, H.~Coffey, A.~Henley, and A.~Sarma,
  ``Supporting code comprehension via annotations: Right information at the
  right time and place,'' in \emph{2020 IEEE Symposium on Visual Languages and
  Human-Centric Computing (VL/HCC)}, 2020, pp. 1--10.

\bibitem{Wettel07Habitability}
R.~Wettel and M.~Lanza, ``Program comprehension through software
  habitability,'' in \emph{15th IEEE International Conference on Program
  Comprehension (ICPC '07)}, 2007, pp. 231--240.

\bibitem{Wettel11ICSE}
R.~Wettel, M.~Lanza, and R.~Robbes, ``Software systems as cities: a controlled
  experiment,'' in \emph{2011 33rd International Conference on Software
  Engineering (ICSE)}, 2011, pp. 551--560.

\bibitem{Tolman1948CMaps}
E.~C. Tolman, ``\BIBforeignlanguage{en}{Cognitive maps in rats and men},''
  \emph{\BIBforeignlanguage{en}{Psychol. Rev.}}, vol.~55, no.~4, pp. 189--208,
  Jul. 1948.

\bibitem{Burgess2002SMemory}
N.~Burgess, E.~A. Maguire, and J.~O'Keefe, ``\BIBforeignlanguage{en}{The human
  hippocampus and spatial and episodic memory},''
  \emph{\BIBforeignlanguage{en}{Neuron}}, vol.~35, no.~4, pp. 625--641, Aug.
  2002.

\bibitem{Cox2005CMap}
\BIBentryALTinterwordspacing
A.~Cox, M.~Fisher, and P.~O'Brien, ``Theoretical considerations on navigating
  codespace with spatial cognition,'' in \emph{Proceedings of the 17th Annual
  Workshop of the Psychology of Programming Interest Group, {PPIG} 2005,
  Brighton, UK, June 29 - July 1, 2005}.\hskip 1em plus 0.5em minus 0.4em\relax
  Psychology of Programming Interest Group, 2005, p.~9. [Online]. Available:
  \url{http://ppig.org/library/paper/theoretical-considerations-navigating-codespace-spatial-cognition}
\BIBentrySTDinterwordspacing

\bibitem{Green1995Imagery}
T.~R.~G. Green and R.~Navarro, ``Programming plans, imagery, and visual
  programming,'' in \emph{{IFIP} Advances in Information and Communication
  Technology}, ser. IFIP advances in information and communication
  technology.\hskip 1em plus 0.5em minus 0.4em\relax Boston, MA: Springer US,
  1995, pp. 139--144.

\bibitem{Jones2007SpatialSkills}
\BIBentryALTinterwordspacing
S.~J. Jones and G.~E. Burnett, ``Spatial skills and navigation of source
  code,'' in \emph{Proceedings of the 12th Annual SIGCSE Conference on
  Innovation and Technology in Computer Science Education}, ser. ITiCSE
  '07.\hskip 1em plus 0.5em minus 0.4em\relax New York, NY, USA: Association
  for Computing Machinery, 2007, p. 231–235. [Online]. Available:
  \url{https://doi.org/10.1145/1268784.1268852}
\BIBentrySTDinterwordspacing

\bibitem{Huang2019fmri}
Y.~Huang, X.~Liu, R.~Krueger, T.~Santander, X.~Hu, K.~Leach, and W.~Weimer,
  ``Distilling neural representations of data structure manipulation using fmri
  and fnirs,'' in \emph{2019 IEEE/ACM 41st International Conference on Software
  Engineering (ICSE)}, 2019, pp. 396--407.

\bibitem{MargulieuxCognitive2019}
\BIBentryALTinterwordspacing
L.~E. Margulieux, ``Spatial encoding strategy theory: The relationship between
  spatial skill and stem achievement,'' in \emph{Proceedings of the 2019 ACM
  Conference on International Computing Education Research}, ser. ICER
  '19.\hskip 1em plus 0.5em minus 0.4em\relax New York, NY, USA: Association
  for Computing Machinery, 2019, p. 81–90. [Online]. Available:
  \url{https://doi.org/10.1145/3291279.3339414}
\BIBentrySTDinterwordspacing

\bibitem{Duff99VisuoSpatialWM}
S.~C. Duff and R.~H. Logie, ``\BIBforeignlanguage{en}{Storage and processing in
  visuo-spatial working memory},'' \emph{\BIBforeignlanguage{en}{Scand. J.
  Psychol.}}, vol.~40, no.~4, pp. 251--259, Dec. 1999.

\bibitem{Furley10CorsiBlockAthletes}
P.~Furley and D.~Memmert, ``Differences in spatial working memory as a function
  of team sports expertise: The corsi block-tapping task in sport psychological
  assessment,'' \emph{Perceptual and Motor Skills}, vol. 110, no.~3, pp.
  801--808, 2010, pMID: 20681333.

\bibitem{Raffaella09Corsi}
R.~Nori, S.~Grandicelli, and F.~Giusberti, ``Individual differences in
  visuo-spatial working memory and real-world wayfinding.'' \emph{Swiss Journal
  of Psychology / Schweizerische Zeitschrift für Psychologie / Revue Suisse de
  Psychologie}, vol.~68, pp. 7--16, 2009, place: Switzerland Publisher: Verlag
  Hans Huber.

\bibitem{Jin18CORSI_UI}
Y.~Jin, N.~Tintarev, and K.~Verbert, ``Effects of individual traits on
  diversity-aware music recommender user interfaces,'' in \emph{Proceedings of
  the 26th Conference on User Modeling, Adaptation and Personalization}, ser.
  UMAP '18.\hskip 1em plus 0.5em minus 0.4em\relax New York, NY, USA:
  Association for Computing Machinery, 2018, p. 291–299.

\bibitem{Huang20VisualRealism}
\BIBentryALTinterwordspacing
J.~Huang and A.~Klippel, ``The effects of visual realism on spatial memory and
  exploration patterns in virtual reality,'' in \emph{Proceedings of the 26th
  ACM Symposium on Virtual Reality Software and Technology}, ser. VRST
  '20.\hskip 1em plus 0.5em minus 0.4em\relax New York, NY, USA: Association
  for Computing Machinery, 2020. [Online]. Available:
  \url{https://doi.org/10.1145/3385956.3418945}
\BIBentrySTDinterwordspacing

\bibitem{Corsi1972HumanMA}
P.~M. Corsi, ``Human memory and the medial temporal region of the brain.''
  1972.

\bibitem{OstroskyDigitSpan2007}
F.~Ostrosky-Sol{\'\i}s and A.~Lozano, ``\BIBforeignlanguage{en}{Digit span:
  Effect of education and culture},'' \emph{\BIBforeignlanguage{en}{Int. J.
  Psychol.}}, vol.~41, no.~5, pp. 333--341, Oct. 2006.

\bibitem{Von_Bastian2013Tatool}
C.~C. von Bastian, A.~Locher, and M.~Ruflin, ``\BIBforeignlanguage{en}{Tatool:
  a java-based open-source programming framework for psychological studies},''
  \emph{\BIBforeignlanguage{en}{Behav. Res. Methods}}, vol.~45, no.~1, pp.
  108--115, Mar. 2013.

\bibitem{Minelli2015InteractionData}
R.~Minelli, A.~Mocci, and M.~Lanza, ``I know what you did last summer - an
  investigation of how developers spend their time,'' in \emph{2015 IEEE 23rd
  International Conference on Program Comprehension}, 2015, pp. 25--35.

\bibitem{RT52}
\BIBentryALTinterwordspacing
A.~T. Welford, ``The psychological refractory period and the timing of
  high-speed performance - a review and a theory,'' \emph{British Journal of
  Psychology.General Section}, vol.~43, no.~1, p.~2, Feb 01 1952, last updated
  - 2013-02-22. [Online]. Available:
  \url{https://www.proquest.com/scholarly-journals/psychological-refractory-period-timing-high-speed/docview/1293556669/se-2}
\BIBentrySTDinterwordspacing

\bibitem{Minelli16TamingIDE}
R.~Minelli, A.~Mocci, R.~Robbes, and M.~Lanza, ``Taming the ide with
  fine-grained interaction data,'' in \emph{2016 IEEE 24th International
  Conference on Program Comprehension (ICPC)}, 2016, pp. 1--10.

\bibitem{Penta:ICPC:2007}
M.~Di~Penta, R.~Stirewalt, and E.~Kraemer, ``Designing your next empirical
  study on program comprehension,'' in \emph{15th IEEE International Conference
  on Program Comprehension (ICPC '07)}, 2007, pp. 281--285.

\bibitem{Lucca:ICPC:2006}
G.~Di~Lucca and M.~Di~Penta, ``Experimental settings in program comprehension:
  Challenges and open issues,'' in \emph{14th IEEE International Conference on
  Program Comprehension (ICPC'06)}, 2006, pp. 229--234.

\bibitem{Wagner:TSE:2021}
S.~Wagner and M.~Wyrich, ``Code comprehension confounders: A study of
  intelligence and personality,'' \emph{IEEE Transactions on Software
  Engineering}, pp. 1--1, 2021.

\end{thebibliography}


\end{document}